\documentclass[]{aa}
\usepackage{hyperref}
\usepackage{comment}
\usepackage[dvipsnames]{xcolor}
\usepackage{soul}
\usepackage{mathtools}
\usepackage{caption}
\usepackage{dblfloatfix}
\usepackage{placeins}
\usepackage{array}  

\usepackage{xcolor}
\hypersetup{    
    colorlinks=true,
    citecolor = blue,
    linkcolor=blue,
    filecolor=magenta,      
    urlcolor=cyan,
}
\usepackage{multirow}
\usepackage{multicol}
\usepackage{pdflscape} 

\usepackage{graphicx}
\usepackage{txfonts}

 \def\mso{\,\mathrm{M_\odot}}
 \def\msun{\mathrm{M_\odot}}

 \def\Msun{\,\mathrm{M_\odot}}

 \def\kms{\, \mathrm{km\,s^{-1}}}

 \def\Yc{Y_\mathrm{c}}
 \def\Ys{Y_\mathrm{s}}
 \def\Mi{M_\mathrm{i}}
 
 \def\Menv{M_\mathrm{env}}

 \def\logTeff{\log \Teff \, \mathrm{[K]}}
 \def\logLLsun{\log L  \, \mathrm{[L_\odot]}}
 \def\Pi{P_\mathrm{i}}
 \def\qi{q_\mathrm{i}}
 \def\Mpi{M_\mathrm{1,i}}
 \def\Teff{T_\mathrm{eff}}

 \def\vini{v_\mathrm{rot, i}}

 \def\Ca{^{12}\mathrm{C}}
 \def\Cb{^{13}\mathrm{C}}
 \def\Ciso{^{12}\mathrm{C}/^{13}\mathrm{C}}
 \def\Oa{^{16}\mathrm{O}}
 \def\Ob{^{17}\mathrm{O}}
 \def\Oiso{^{16}\mathrm{O}/^{17}\mathrm{O}}
 
 \def\Menv{M_\mathrm{env}}
 
 \def\logNNi{\log \mathrm{N/N_{i}}}
 
 \def\logNC{\log \mathrm{(N/C)}}

\def\simle{\mathrel{\hbox{\rlap{\hbox{\lower4pt\hbox{$\sim$}}}\hbox{$<$}}}}
 \def\simgr{\mathrel{\hbox{\rlap{\hbox{\lower4pt\hbox{$\sim$}}}\hbox{$>$}}}}

\newcommand{\Fig}[1]{Fig.\,\ref{#1}}
\newcommand{\Figure}[1]{Figure\,\ref{#1}}

\begin{document}

   \title {A comprehensive grid of massive binary evolution models for the Galaxy -- Surface properties of post-mass transfer stars}
    

   \author{Harim Jin \inst{1}
          \and Norbert Langer \inst{1,2} 
          \and Andrea Ercolino \inst{1} 
          \and Selma E. de Mink \inst{3} 
          }

   \institute{Argelander Institut für Astronomie,
              Auf dem Hügel 71, DE-53121 Bonn, Germany\\
              \email{hjin@astro.uni-bonn.de}
         \and
          Max-Planck-Institut für Radioastronomie, Auf dem Hügel 69, DE-53121 Bonn, Germany
         \and
          Max Planck Institute for Astrophysics, Karl-Schwarzschild-Straße 1, 85748 Garching bei München, Germany
    }

   \date{Received Month Day, 2025; accepted Month Day, 2025}

 
  \abstract
   {Massive stars often evolve in binary systems, in which binary interactions significantly affect their evolution. Massive stars in the Galaxy serve as valuable testbeds for this due to their proximity.}
   {We computed the evolution of more than 38\,000 galactic binary systems with initial primary star masses of 5$\dots$100\,M$_{\odot}$. 
   In this paper, we aim to investigate the surface properties of post-mass transfer mass  donor and mass gainer stars through core hydrogen burning, core helium burning, and for the  pre-supernova stage.}
   {The models are computed with MESA, incorporating detailed stellar and binary physics, including internal differential rotation, magnetic angular momentum transport, mass-dependent overshooting, stellar wind mass-loss, mass and angular momentum transfer and tidal interaction. They incorporate a new extensive nuclear network for hydrogen burning, which allows us to track the full range of hydrogen burning nucleosynthesis products, from the light elements to aluminum. The widest, non-interacting binary models in our grid effectively serve as single star models.}
   {We find that mass gainers and donors may evolve through long-lived blue and yellow supergiant stages during core helium burning where single stars of the same mass remain red supergiants. Furthermore, some of our mass gainers evolve into more luminous yellow and blue supergiants prior to core collapse than single stars, while some mass donors end their life as red or yellow supergiants, showing a rich diversity in supernova progenitors. We show that the surface elemental and isotopic abundances carry valuable information about a star's evolutionary history and can be used to distinguish binary interaction products from single stars. 
   }
   {Our binary model grid may serve as a tool for identifying post-mass transfer stars and supernovae, and holds potential for population studies, supernova modeling, and guidance of future observations.}

   \maketitle
%

\section{Introduction}\label{sec:intro}
Most massive stars are born in binary systems and interact with their companions throughout their evolution \citep{Sana2012,deMink2014}. Binary interactions affect chemical \citep[e.g.,][]{Farmer2023}, mechanical \citep[e.g.,][]{Wagg2025}, and radiative \citep[e.g.,][]{Goetberg2020} feedback of massive stars to their surroundings, shaping the evolution of star-forming galaxies across cosmic time. Products of binary interactions give rise to X-ray binaries, peculiar supernovae, and magnetic stars \citep[see][for a review]{Langer2012}. Their remnants can become gravitational wave sources, which may provide an independent determination of the present-day expansion rate of the Universe \citep{Abbott2017}. However, the key physical processes governing the evolution of massive binaries remain poorly understood.

Many detailed binary evolution models have been computed to be compared with observations. Much attention has been given to predicting the final fates of massive stars \citep[e.g., supernovae and gravitational wave sources as in][]{Marchant2016,Eldridge2018,Laplace2020,Marchant2021} or their late evolutionary stages \citep[e.g., Be X-ray binaries as in][]{Shao2015,Rocha2024,Ge2024}. In contrast, less focus has been placed on earlier evolutionary stages, such as main sequence stars in binaries (see however early work by \citet{Nelson2001,deMink2007,Sen2022}). Since the main sequence phase accounts for approximately 90\% of a massive star's lifetime, it is observationally favorable for using stars on the main sequence to understand binary interaction history. 

Recent large observational campaigns, such as the VLT-FLAMES Tarantula survey (VFTS) for the Large Magellanic Cloud \citep{Evans2011}, the IACOB spectroscopic survey for the Galaxy \citep{SimonDiaz2014}, and the Binarity at LOw Metallicity (BLOeM) campaign for the Small Magellanic Cloud \citep{Shenar2024}, have provided high-quality photometric and spectroscopic data for hundreds of early type stars. These datasets enable statistically significant and homogeneous samples of massive star parameters. 
Synthetic binary populations predict most of the mass gainers and merger products to be main sequence stars or supergiants \citep[e.g.,][]{Vanbeveren2013, Justham2014, deMink2014, Langer2020}.
However, identifying which stars have undergone binary interactions is not always straightforward.

Surface abundances can be one avenue for identifying binary interaction products. Previous studies have investigated the surface abundances of binary models regarding the effects of tidally-induced mixing \citep{deMink2009tides,Song2013,Song2018tides}, mergers \citep{Glebbeek2013,Schneider2019,Schneider2020,Menon2024}, and mass transfer \citep{Wellstein1999,Wellstein2001,Petrovic2005,Langer2008,deMink2009mt,Langer2010,Langer2012,Song2018mt,Renzo2021,Pauli2022,Sen2022,Klencki2022,Sen2023,Xu2025}. Particularly, surface abundance predictions for post-mass transfer stars so far have been limited to subsets of interacting binaries and/or limited to helium and CNO abundances.

This work aims to provide highly needed predictions for post-mass transfer stars which are comprehensive in terms of binary parameter space and complete in terms of hydrogen-burning nucleosynthesis, the products of which span from hydrogen to aluminum.
In this study, we introduce a new binary evolution model grid that tracks the detailed surface abundance evolution in both stellar components. We provide surface abundances predictions for main sequence stars and post-main sequence stars and their locations in the Hertzsprung-Russell diagram (HRD). In Section~\ref{sec_mod}, we describe the details of the numerical method and physics assumptions adopted for our models, and give an overview of our model grid. In Section~\ref{sec_res}, we present our main predictions regarding the evolution in the HRD and surface abundances. In Section~\ref{sec_dis}, we compare our model grid with other model grids and address the limitations of our models. Finally, we summarize and conclude our findings in Section~\ref{sec_con}.

\section{Methods}\label{sec_mod}
We use the one-dimensional stellar evolution code MESA \citep[version 10398;][]{Paxton2011,Paxton2013,Paxton2015,Paxton2018,Paxton2019} to compute binary evolution models. The input files, including initial abundances, opacity tables, nuclear networks, and zero-age main sequence (ZAMS) models, necessary to reproduce the MESA calculations, are available online along with the complete binary model grid\footnote{\url{https://wwwmpa.mpa-garching.mpg.de/stellgrid/}}.

\subsection{Stellar physics}\label{sec_physics}
The evolution of both stars in a binary system is followed simultaneously, starting from the ZAMS. We adopt the same physics assumptions as in \citet{Jin2024} for individual stellar components in binary systems. Below, we briefly summarize the key aspects.

We adopt the proto-solar abundances of \citet{Asplund2021} as the initial abundances. Opacity tables are tailored to these abundances, with high-temperature ($\logTeff > 4$) opacities generated via the OPAL project website \citep{Iglesias1996} and low-temperature ($\logTeff < 4$) opacities taken from \citet{Ferguson2005}. The nuclear network is the one used in STERN \citep{Heger2000,Jin2024}, which includes detailed reactions of the pp-chains, the CNO-cycles, the Ne-Na cycle, the Mg-Al cycle, as well as helium, carbon, neon, and oxygen burning. The network tracks the abundances of all stable isotopes up to silicon, along with radioactive $^{26}\rm Al$. We follow the surface abundances of
$^{1,2}\rm H$, $^{3,4}\rm He$, $^{6,7}\rm Li$, $^{9}\rm Be$, $^{10,11}\rm B$,
$^{12,13}\rm C$, $^{14,15}\rm N$, $^{16-18}\rm O$, $^{19}\rm F$, $^{20-22}\rm Ne$,
$^{23}\rm Na$, $^{24-26}\rm Mg$, $^{26}\rm Al_g$, $^{27}\rm Al$, $^{28,29}\rm Si$,
$^{40}\rm Ca$, and $^{56}\rm Fe$. Calcium and iron are not included in the nuclear network; instead, calcium represents the mass fraction of elements not explicitly included, and iron is used to scale wind mass-loss rates. We adopt the reaction rates from the JINA REACLIB \citep{Cyburt2010}.

For convection, we adopt the standard mixing length theory \citep{Bohm1958} with a mixing length parameter of $\alpha_\mathrm{MLT}=1.5$ \citep[see e.g., the discussion in][]{Maeder1989}, using the Ledoux criterion for convection and the semiconvection with an efficiency of $\alpha_\mathrm{sc}=1$ \citep{Langer1983}. For the post-main sequence evolution, we switch to MLT++ to avoid convergence errors \citep{Paxton2013}. We implement mass-dependent overshooting \citep{Hastings2021}, where $\alpha_\mathrm{ov}$ increases linearly from 0.1 at 1.66 $\mso$ \citep{Claret2016} to 0.3 at 20 $\mso$ \citep{Brott2011}, matching observational constraints. Thermohaline mixing is treated following \citet{Cantiello2010} with $\alpha_\mathrm{th}=1$ \citep{Kippenhahn1980}.

The stellar wind mass-loss prescription combines the rates of \citet{Vink2001} for hot, hydrogen-rich stars ($X>0.7,\; \Teff \gtrsim 25, \mathrm{kK}$) and the maximum of this and the rate of \citet{Nieuwenhuijzen1990} for cool, hydrogen-rich stars ($X>0.7,\; \Teff \lesssim 25, \mathrm{kK}$). For stars with $X<0.7$, we use mass-loss rates for Wolf-Rayet stars from \citet{Nugis2000} and \citet{Yoon2017}, following \citet{Pauli2022}. Rotationally-enhanced mass-loss is implemented following \citet{Heger2000,Yoon2005}.

The initial rotation rate of our models is set to 20\% of the critical rotation, corresponding to the lower peak in the bimodal distribution of rotational velocities from \citet[][see their fig.\,8]{Dufton2013}. Note that this is based on the distribution of early B-type stars, and stars with earlier types \citet[e.g.,][]{RamirezAgudelo2013} and later types \citet[e.g.,][]{Zorec2012} seem to have different distributions. We discuss how this choice affects our results in Sect.~\ref{sec_lim}. Rotational mixing is implemented as in \citet{Heger2000}, with a diffusion coefficient factor of $f_\mathrm{c}=1/30$ \citep{Chaboyer1992} and a parameter for the inhibiting effect of stabilizing chemical gradients set to $f_\mathrm{\mu}=0.1$ \citep{Yoon2006}. The Spruit-Tayler dynamo and its associated angular momentum transport are implemented as in \citet{Petrovic2005} with $f_\mathrm{\nu}=1$.

\subsection{Binary physics}
We adopt the same binary physics as in \citet{Marchant2017}, except for the mass transfer scheme. When the primary star (initially the more massive component) is undergoing core hydrogen burning, we use a scheme in which the mass transfer rate is determined implicitly to ensure that the primary remains within its Roche lobe in semi-detached systems. In contact systems, the rate is calculated such that both stars share the same equipotential surface. Once the primary depletes its core hydrogen, we switch to an explicit scheme following \citet{Kolb1990}, which accounts for the optically thin region above the photosphere. This approach is supported by recent 3D simulations, which showed quantitatively similar mass transfer rates \citep{Ryu2025}. This is particularly relevant for giants and supergiants, where the pressure scale height can be comparable to their stellar radius.

The efficiency of mass transfer (the fraction of transferred mass that is accreted by the companion) is regulated by the accretor's rotation \citep{Petrovic2005}. As the accretor gains angular momentum \citep{deMink2013}, it can spin up to near-critical rotation, even with the accretion of just a few percent of its initial mass \citep{Packet1981,Ghodla2024}. If this happens, accretion is halted, and any additional transferred material is assumed to be lost from the system, carrying away the accretor’s specific orbital angular momentum. In our framework, the mass transfer efficiency is not treated as a free parameter but is instead computed self-consistently based on the accretor’s and the orbit’s response. We discuss the limitations of this approach in Sect.~\ref{sec_lim}.

We assume circular binary orbits, evolving them simultaneously with the two stellar components. Orbital angular momentum changes due to stellar wind mass loss, tidal interactions, and mass exchange between the two stars in the system. Tidal effects are implemented following \citet{Detmers2008}, with the tidal timescale based on \citet{Zahn1977}. We note that tidal interactions in stars with radiative envelopes remain uncertain, and tides can alter stellar structure, spin, and orbital evolution in ways that are not fully captured by our current setup \citep[e.g.,][]{Witte1999,Fuller2017,Sciarini2024}.

In the case of critically rotating accretors, we assume that material is lost in the vicinity of the accretor. However, whether sufficient power is available for this process remains unclear. The power source could be the radiation from the stars, and we assess whether the radiative power is sufficient to expel the non-accreted material to infinity. We consider two scenarios: 1) material is lost at the surface of the accretor (which we will refer to as the Lower $\dot{M}$ limit) or at the Roche lobe surface of the accretor (Upper $\dot{M}$ limit). As it is energetically harder to expel material from the surface of the accretor than of its Roche lobe, the mass transfer rate corresponding to the Lower $\dot{M}$ limit is smaller than the Upper $\dot{M}$ limit. While we do not terminate models when these limits are exceeded, we flag them for further analysis. For details on our termination conditions and comparisons with those used in other binary model grids, see the Appendix~\ref{app_term}.

\subsection{Binary model grid}\label{sec_grid}
Our binary model grid consists of $38\,430$ evolutionary sequences. The ranges of initial orbital periods are chosen to span the full range of systems that interact. The grid include systems undergoing Roche lobe overflow already at ZAMS at the shortest-period end, and those that do not experience Roche lobe overflow at the longest-period end. The grid spans initial primary masses ($\Mpi$) of $5-100\mso$ ($\log \Mpi \,[\msun]=0.70-2.00$) with a step size of $\Delta \log \Mpi \,[\msun]=0.05$, initial mass ratios ($\qi$) of $0.1-0.95$ with a step size of $\Delta \, \qi = 0.05$, and initial orbital periods ($\Pi$) of $0.3 - 5011\, \rm d$ for $\log \Mpi \,[\msun] \leq 1.50$ and $1.3 - 3162\, \rm d$ otherwise with a step size of $\Delta\log \Pi \mathrm{[d]}=0.05$. 

\begin{figure}
	\centering
	\includegraphics[width=\linewidth]{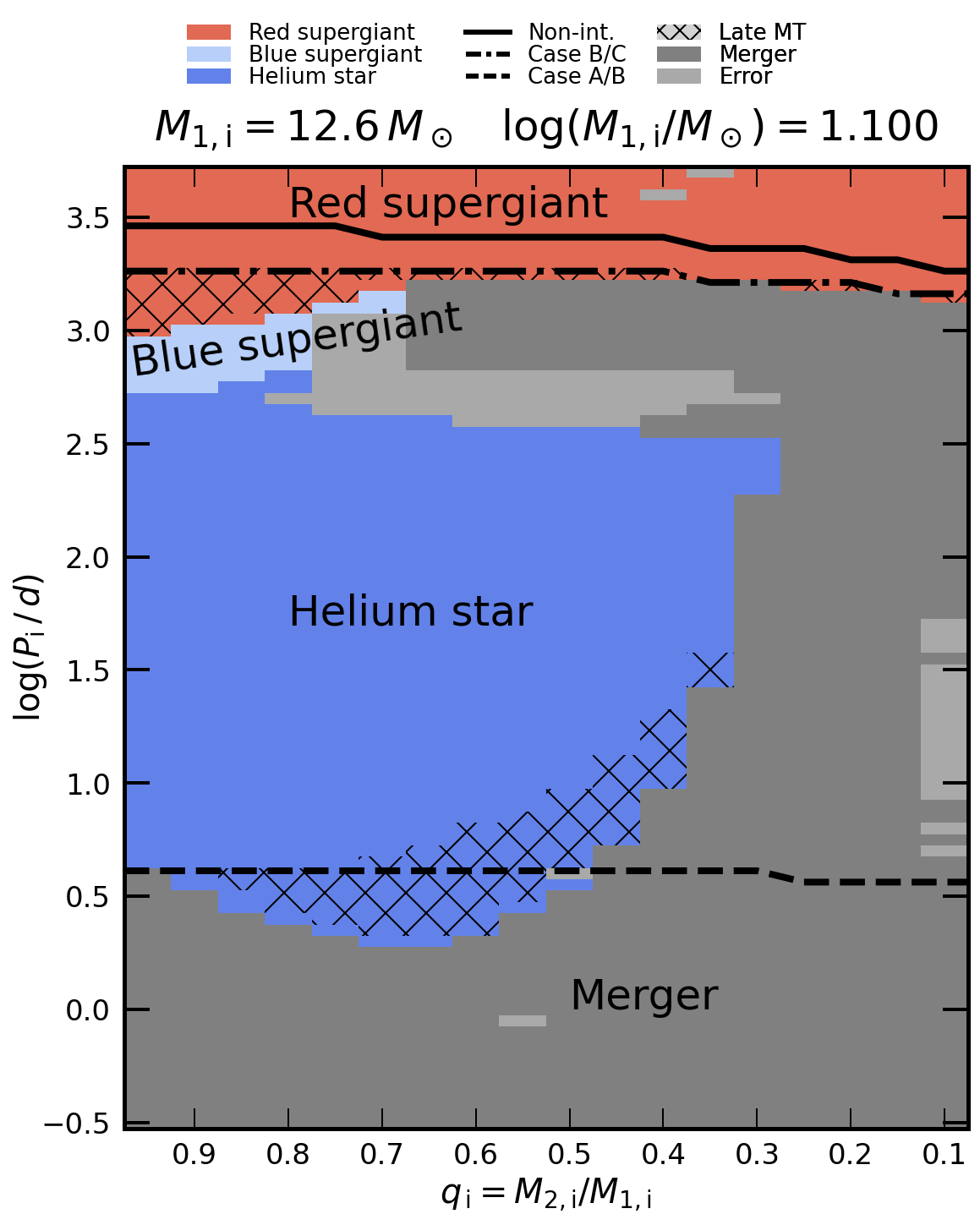}
	\caption{An overview of our binary evolution models with an initial primary mass of $12.6\mso$. Each pixel represents one detailed binary evolution sequence, with colors indicating the status of the primary star in the midpoint of core helium burning (when central helium mass fraction is $\Yc=0.5$). At this point, the secondary is a core hydrogen burning star. Black lines delineate the boundaries between different mass transfer cases, while the black-hatched region highlights models that will undergo another phase of mass transfer after primary's core helium depletion ("Late MT"). Models that are expected to merge, and models that did not converge due to unknown error are also indicated.}
	\label{fig_Pq}
\end{figure}

\Figure{fig_Pq} illustrates the main outcomes of the 1530 models with an initial primary mass of $12.6\mso$. Depending on the initial mass ratio and orbital period, binary evolution models exhibit different mass transfer histories and termination conditions. For example, systems undergo Case A, Case B \citep{Kippenhahn1967}, and Case C \citep{Lauterborn1970} mass transfer with increasing initial orbital period, and do not interact through Roche lobe overflow for the highest initial orbital period. Some Case B systems undergo a second phase of mass transfer after core helium depletion, potentially leading to interacting supernovae \citep{Laplace2020,Wu2022,Ercolino2024,Ercolino2025}. Similarly, some Case A systems undergo additional mass transfer phases. A complete list of such overview plots for the other initial primary masses in our grid is provided in Appendix~\ref{app_Pq}. 

Although the model set up has been designed to reach core carbon depletion \citep[cf.][]{Jin2024}, many do not reach this stage due to numerical issues or the onset of unstable mass-transfer. We provide a detailed overview of these termination points in Appendix\,\ref{app_term}.

\subsection{Our approach}\label{sec_app}
In this paper, we investigate the areas populated by single stars and post-mass transfer stars in the HRD and surface abundance planes. We check the areas that are populated by only one of the two groups and identify distinctive features that post-mass transfer stars show, which can be used to differentiate them from their single-star counterparts. For single stars, we use the models from \citet{Jin2024} for initial mass between $5-40\mso$, and extend the mass range up to $100\mso$ and use the extended models as well. The initial rotational velocities of single stars span $0-600 \kms$, but in this paper, we only consider models with initial rotational velocities below 400$\kms$. We do not perform population synthesis calculations to estimate the occurrence rates of such systems, and this will be done in our future work. 

\section{Results}\label{sec_res}
Mass transfer can lead to observable differences in donors and accretors, distinguishing them from single stars. In this section, we present the properties of post-mass transfer stars, with a particular focus on main sequence stars, blue supergiants (BSGs), and red supergiants (RSGs), and their evolution in the HRD and their surface abundances. We do not provide predictions for merger products (see Sect.~\ref{sec_lim} for a discussion).

\subsection{Evolution in the HRD}\label{sec_res1}
In this section, we will first explore how binary stars evolve in the HRD (Section~\ref{sec_tracks}), and provide an overview of model predictions across the whole model grid—for core hydrogen burning stars (Section~\ref{sec_CHB}), core helium burning stars (Section~\ref{sec_CHeB}), and supernova progenitors (Section~\ref{sec_preSN}).

\begin{figure*}
	\centering
	\includegraphics[width=0.97\linewidth]{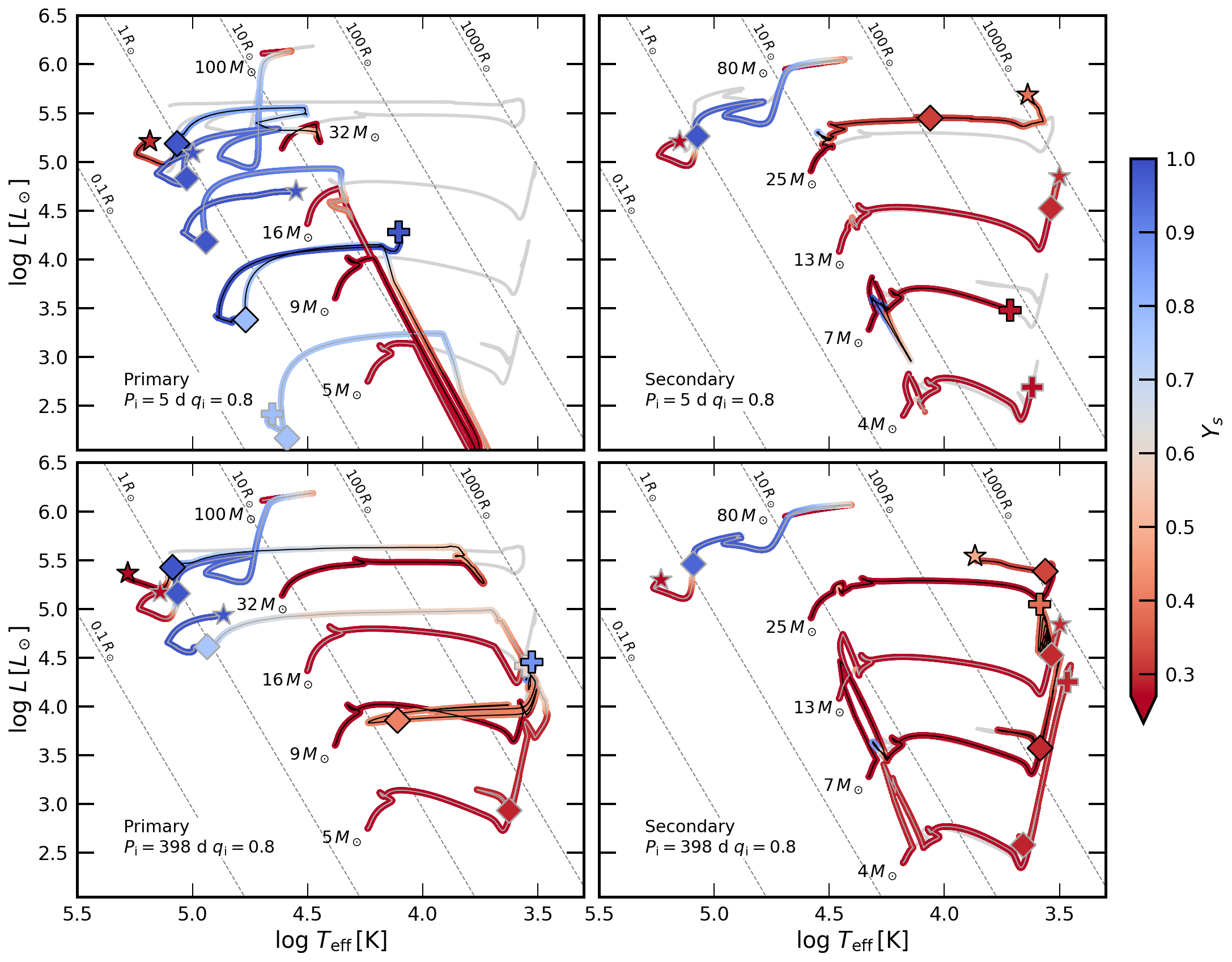}
	\caption{Evolutionary tracks of primaries (left) and corresponding secondaries (right) for selected binary systems with $\qi=0.8$, $\Pi = 5\, \rm d$ (top) and $\Pi = 398\, \rm d$ (bottom) in the Hertzsprung-Russell diagram. The initial masses of the stars are indicated. The color-coding represents the surface helium mass fraction. Diamonds mark the midpoint of core helium burning ($\Yc=0.5$), stars indicate the point of central carbon depletion, and plus signs mark the termination point due to the expected merger. For reference, single star evolutionary tracks are shown as grey lines, plotted up to core helium depletion.}
	\label{fig_tracks}
\end{figure*}

\begin{figure*}
	\centering
	\includegraphics[width=0.95\linewidth]{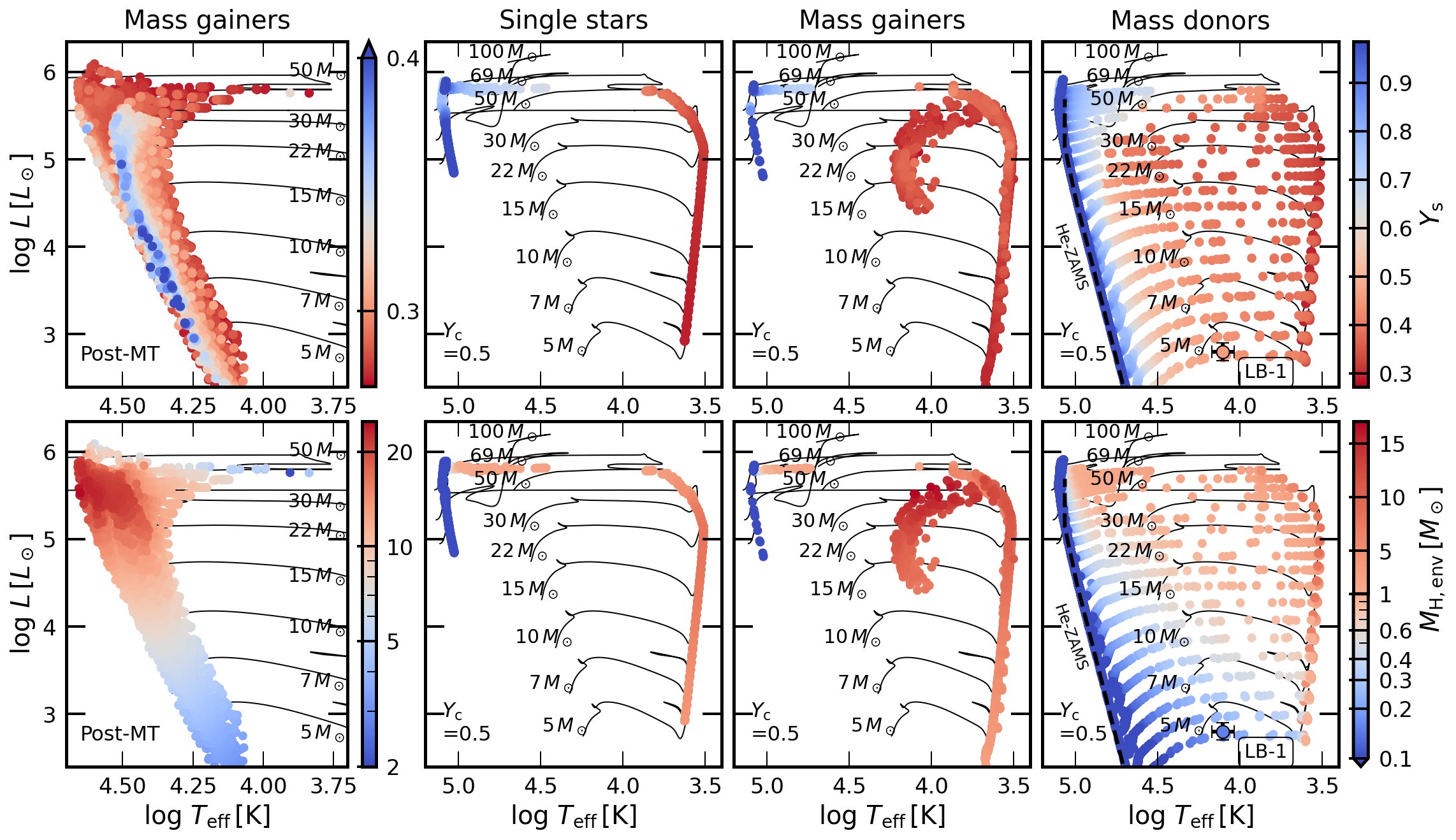}
	\caption{Core hydrogen burning mass gainers after thermal relaxation (first column; see text), and of core helium burning single stars (second), mass gainers (third), and mass donors (fourth) in the Hertzsprung-Russell diagram, each shown at the midpoint of core helium burning ($\Yc=0.5$). Thus, each data point represents one model sequence at a specific evolutionary stage. The color-coding represents surface helium mass fraction (top) and current hydrogen envelope mass (bottom), which is defined as the total mass subtracted by the helium core mass. Black lines represent evolutionary tracks for non-rotating single star models with their initial mass indicated. In the panels in the fourth column contain the data point for stripped star component in the black-hole binary imposter LB-1 \citep{Shenar2020,Schuermann2022}. Note that the colorbar range in the first column differs from that in the remaining columns.}
	\label{fig_HRD3}
\end{figure*}

\begin{figure}
	\centering
	\includegraphics[width=0.97\linewidth]{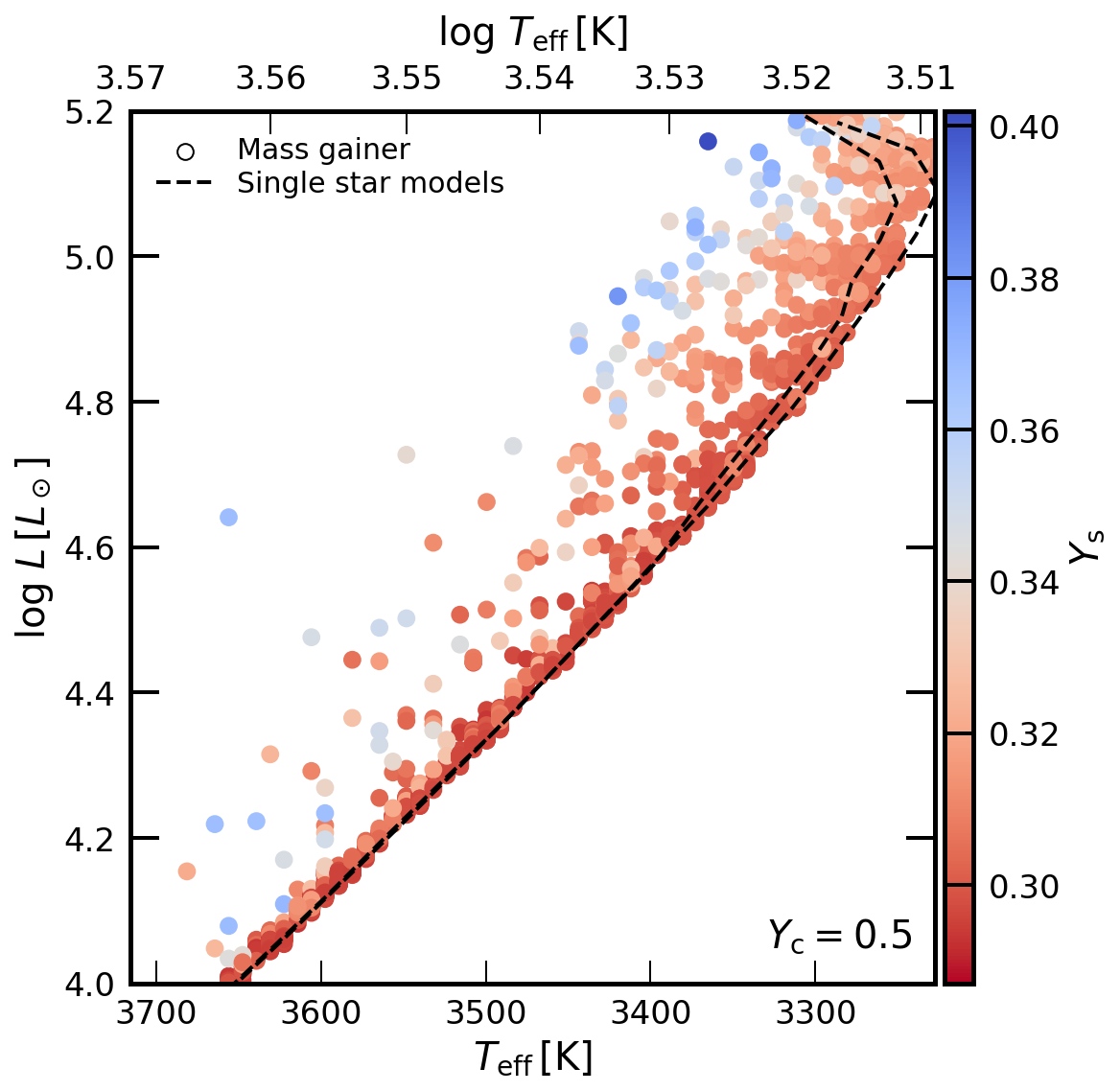}
	\caption{Mass gainer models at the midpoint of core helium burning ($\Yc=0.5$) in the red supergiant region of the Hertzsprung-Russell diagram. The color-coding represents the surface helium mass fraction. Dashed lines indicate the region predicted by single star models.}
	\label{fig_HeTeff}
\end{figure}

\begin{figure*}
	\centering
    \includegraphics[width=0.95\linewidth]{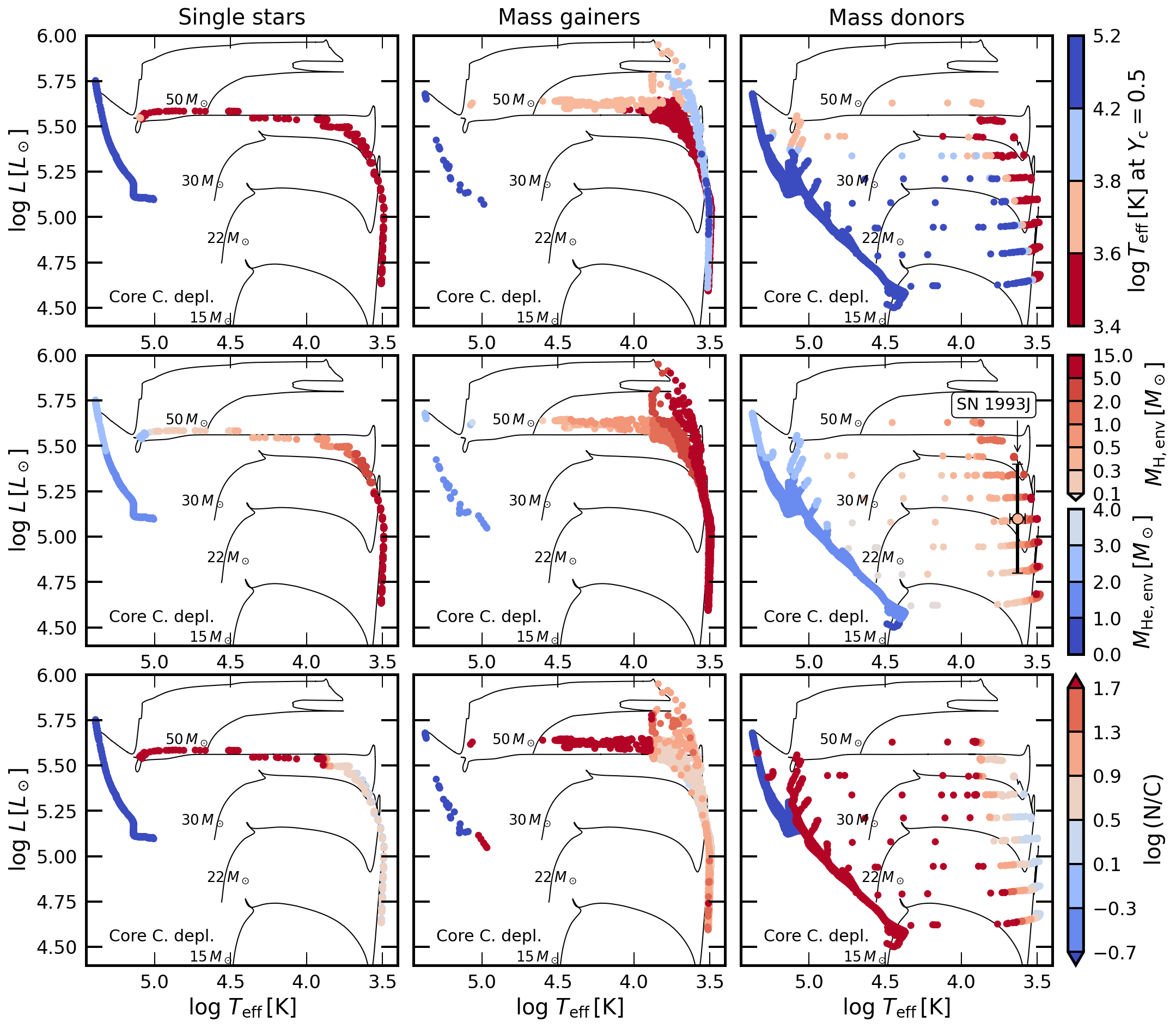}
	\caption{Single stars (left), mass gainers (middle), and mass donors (right) in the Hertzsprung-Russell diagram at core carbon depletion, which well represents the pre-supernova stage. The color-coding represents the effective temperature at the midpoint of core helium burning ($\Yc=0.5$), the remaining hydrogen and helium envelope mass, and nitrogen-to-carbon ratio, from top to bottom. Data points are stacked such that those with higher $\log \Teff$, $\Menv$, $\logNC$ appear above those with lower values. Evolutionary tracks for non-rotating single star models are also presented with their initial mass indicated. In the mass donor panel on the second row, the data point for SN\,1993J progenitor is also shown \citep{Maund2004}.
    }
	\label{fig_preSN}
\end{figure*}

\subsubsection{Exemplary evolutionary tracks}\label{sec_tracks}
Examples of evolutionary tracks for primaries and secondaries are shown in \Fig{fig_tracks} \citep[see also fig.\,2 of][for a similar plot]{Pauli2022}. The primaries and secondaries exhibit different tracks compared to single stars. For example, the secondaries (except for the $80\mso$ one) accrete mass during their main sequence evolution, and appear overluminous during the accretion phase. Many primaries do not reach the RSG dimension due to mass transfer. 

The mass transfer history differs significantly depending on the initial orbital periods (separations). For instance, a 9$\mso$ primary in a 5$\, \rm d$ orbit undergoes nearly complete hydrogen envelope stripping, resulting in a high surface helium mass fraction\footnote{Throughout this paper, we denote the helium mass fractions at the surface and at the center as $\Ys$ and $\Yc$, respectively. Element and isotope names refer to their surface number fractions). For example, $\log \mathrm{(N/C)}$ represents the (base 10) logarithmic surface number ratio of nitrogen to carbon.} at the end of mass transfer. Then it undergoes core helium burning as a stripped helium star. In contrast, the same 9$\mso$ primary in a 398$\, \rm d$ orbit experiences partial envelope stripping, leaving it with a relatively low surface helium mass fraction, and the star undergoes core helium burning near the main sequence band.

Secondaries also show distinct behaviors. While the 25$\mso$ secondary in a 398$\, \rm d$ orbit is a RSG at the midpoint of core helium burning, the same-mass secondary in a 5$\, \rm d$ orbit is a BSG. The two corresponding binary systems experience Case B and Case A mass transfer, respectively. In our models, secondaries in Case A systems can accrete a significant amount of mass because stronger tides in close orbits can prevent accretors from reaching critical rotation, which inhibits efficient mass accretion. When core growth is not too pronounced upon mass transfer—i.e., when the envelope-to-core mass ratio is higher than that of a single star with the same initial mass—the accretor can undergo core helium burning as a BSG \citep{Braun1995}. The lowest-mass secondaries, with initial masses of $4\mso$ and $7\mso$, transfer mass onto the primaries after the main sequence and before core helium burning ($\Pi=5\,\rm d$) or after core helium burning ($\Pi=398\,\rm d$), and are expected to merge.

\subsubsection{Core hydrogen burning stars}\label{sec_CHB}
The first column of \Fig{fig_HRD3} presents the locations, surface helium mass fractions, and hydrogen envelope masses of mass gainers at the moment when the star has thermally relaxed—that is, after more than a thermal timescale has passed since the completion of mass transfer
From this point onward, stars evolve on the main sequence band and their surface helium abundances remain largely unchanged, except for the most luminous ones, which experience strong wind mass loss and become stripped. At this time, mass gainers have companions, which are stripped either fully or partially. 

Strongly helium-enriched mass gainers are located near the ZAMS line, and these stars are secondaries from low initial mass ratio. Initially less massive secondaries have smaller envelope masses, so they experience a less strong dilution than more massive ones. They also have evolved less when mass transfer happened due to their longer core hydrogen burning timescale. These stars will remain helium-enriched throughout the main sequence evolution. Not many helium-rich stars are at the high-luminosity end ($\logLLsun \gtrsim 5$), near the ZAMS line. This is because the highest mass binaries only undergo fast Case A mass transfer, due to strong stellar wind preventing the donors from expanding \citep{Vanbeveren1998,Sen2023}, and do not transfer much helium-rich matter. Fast Case A mass transfer can happen at the early stage of core hydrogen burning of the primary, so there are mass gainers near the ZAMS lines. The hydrogen envelope mass increases with luminosity, except for the most luminous stars, where strong wind mass-loss has reduced the envelop mass. 

\subsubsection{Core helium burning stars}\label{sec_CHeB}
\Figure{fig_HRD3} also compares the locations of single star models, mass gainer models, and mass donor models, each shown at the midpoint of core helium burning. For more than half of the core helium burning phase, the majority of the models ($\gtrsim 70\%$) remain within $\pm 0.1\,\rm dex$, $\pm 0.2\,\rm dex$ with respect to their $\logTeff$, $\logLLsun$ at the midpoint of core helium burning, respectively, except for the most luminous ones ($\logLLsun > 5.7$). The three groups exhibit distinct patterns in the HRD. Single star models are generally located along the Hayashi line (at $\logTeff \sim 3.6$, where the fully convective stars will be located) and have relatively low surface helium abundance ($\Ys \lesssim 0.5$), except for the most massive stars ($\Mi \gtrsim 50 \mso$), which lose mass through strong stellar winds and evolve into Wolf-Rayet stars with high surface helium abundance ($\Ys \gtrsim 0.7$). The most massive mass gainers and mass donors also become Wolf-Rayet stars due to strong winds, but the less massive stars display different behaviors.

Mass gainers, like single stars, can be found along the Hayashi line or have become Wolf-Rayet stars. Notably, many mass gainers also populate the region of $3.7 \lesssim \logTeff \lesssim 4.0$. Only mass gainers that accreted a significant amount of mass through Case A mass transfer end up in that region, such as the $25 \mso$ secondary in a $5 \, \rm d$ orbit shown in \Fig{fig_tracks}. These stars show relatively low surface helium abundance. Most mass gainers, however, accrete little mass (since the majority of systems are Case B systems) and thus behave similarly to single stars, remaining on the Hayashi line. By the time mass gainers reach the midpoint of core helium burning, their companions have already evolved into compact objects in almost all the cases (99.7\%). At this stage, the binary system has either been disrupted by the supernova kick or hosts a compact object. If binary disruption happens, mass gainers are likely to be walkaway/runaway stars \citep{Eldridge2011,Renzo2019,Schuermann2025}.

Mass donors exhibit a wide range of effective temperatures and surface helium abundances, depending on the extent of envelope stripping. The least stripped mass donors lie on the Hayashi line, like single stars, while the most stripped ones are located on the helium-ZAMS line. Unlike single stars, where significant stripping occurs only for initially high-mass stars due to strong stellar winds, stripping in binaries due to mass transfer is independent of initial mass. In most cases, mass donors lose the majority of their envelopes (lie on the helium-ZAMS line). Notably, some mass donors are only partially stripped ($7.3\%$ of all mass donors are more than $0.2\, \rm dex$ away from the helium-ZAMS line in $\logTeff$ in \Fig{fig_HRD3}), with remaining envelope masses of up to several solar masses. In contrast to the nearly completely stripped stars, which quickly evolve into subdwarfs, helium stars, or Wolf-Rayet stars, the partially stripped stars remain in the blue, yellow, or red supergiant regime for most of their core helium-burning phase. Partially-stripped donors are overluminous (i.e., stars that appear more luminous than expected for their mass and evolutionary stage). The degree of stripping and the surface helium abundance depend on the initial orbital period. 

All the mass donors at this time have a main sequence companion. In most cases, donors are completely stripped and their main sequence companions will likely outshine the donors in the optical \citep{Goetberg2017}. However, if mass donors undergo partial stripping and remain as supergiants, then their brightness in the optical can be comparable or higher than their main sequence companions. Main sequence companions can be more massive than the supergiants due to mass transfer. It is interesting that some low-mass RSGs in the Small Magellanic Cloud were found to have more massive companions \citep{Patrick2022}, and our models with partially stripped donors may explain them.

LB-1 is a galactic binary system which was once thought to consist of a massive black hole and a B-type main sequence star. However, recent studies showed that the system does not host a black hole, but a stripped B-type star. Many works investigated the channel that the stripped star is on the fast contraction phase right after mass transfer. Our mass donor models show that it can be actually undergoing core helium burning, rather than being in a transitional phase. In \Fig{fig_HRD3}, we present the observational results of the stripped star in LB-1. Its effective temperature, luminosity, and surface helium mass fraction \citep{Shenar2020} well match our predictions from mass donors with initial mass of $5.0\Msun$. 
The remaining hydrogen envelope mass, estimated by \citet{Schuermann2022} to be approximately $0.15 \Msun$, is slightly lower than in our models ($\sim 0.20 \Msun$), as their mass donors are more stripped and in the contraction phase. Since our partially stripped mass donors can remain in the region of the HRD that was previously thought to be populated by thermally contracting stars during core helium burning phase, this partial-stripping channel increases the predicted number of systems like LB-1. Here, we compare our models only with LB-1, which is a prototype of this kind. We note that there are more LB-1–like systems; for an updated list, readers may refer to, e.g., fig.\,5 of \citet{Marchant2024}.

\Figure{fig_HeTeff} compares the effective temperatures of mass gainer models and single star models in the RSG region. While single star models show a narrow range of effective temperatures at each luminosity, mass gainer models exhibit a wider range. This difference arises from the helium enrichment in the envelopes of mass gainers. As opacity primarily comes from the $\mathrm{H}^{-}$ ion for $\Teff < 5000 \, \rm K$, a lower hydrogen abundance at the surface leads to a reduction in opacity, which shifts the Hayashi line leftward in the HRD. (The highest surface helium mass fraction achieved in our RSG single star models presented in \Fig{fig_HeTeff} is $\Ys = 0.32$.) This implies that if a group of RSG stars from the same environment (e.g., clusters) shows a few with higher effective temperatures than the majority, it suggests these stars may have undergone binary interactions that resulted in helium enrichment in their envelopes. Similarly, binary effects can cause a luminosity spread  of red supergiants in coeval young star clusters \citep{Wang2025}.

\subsubsection{Immediate supernova progenitors}\label{sec_preSN}

\Figure{fig_preSN} shows the locations of single stars, mass gainers, and mass donors at core carbon depletion in the HR diagram. As the remaining lifetime until core-collapse is very short, their stellar structure at this time can be a proxy of their pre-supernova structure. In a few models, late mass transfer events can alter pre-supernova structure \citep[e.g.,][]{Wu2022, Ercolino2024, Ercolino2025}. Only models with carbon core mass of $\simgr 1.435 \mso$, which are expected to undergo core-collapse, are presented in the figure \citep{Nomoto1984, Xu2025}. Single star models are located in two branches: one branch hotter than $\logTeff \sim 5.0$ consisting of stars that already lost most of their envelope and were hot by the time of core helium burning (Wolf-Rayet stars; see \Fig{fig_HRD3}) and the other branch cooler than that consisting of stars that were mostly on the RSG branch then. Stars in the hotter branch lost their entire hydrogen envelope and reveal their bare helium cores. On the other hand, stars in the cooler branch have different amounts of hydrogen envelope, and envelope mass decreases for higher luminosity.

Mass gainers and mass donors also display both the hotter branch and the cooler branch. For mass donors, the hotter branch reaches a far lower luminosity of $\logLLsun \sim 4.5$, which contains the relics of stripped helium stars (\Fig{fig_HRD3}). For the cooler side, there are many mass gainers which end their lives as RSGs. Those are the stars that underwent core helium burning as blue or yellow supergiants so they could avoid strong RSG mass loss for a considerable time. They do not show a monotonic decrease of the hydrogen envelope mass with luminosity. Mass donors show a wide range of temperatures, due to the different degrees of stripping. When these stars explode as supernovae, their photometric and spectroscopic characteristics differ from those of normal Type IIP supernovae \citep{Dessart2024}. For example, partially stripped stars can explode as Type IIb or IIL supernovae. In \Fig{fig_preSN}, as an exemplary case, we show the Type IIb supernova SN\,1993J, whose envelope mass is estimated to be $\sim 0.3 \Msun$ \citep{Maund2004}, and this can be well explained by our partially stripped mass donors. Fully stripped donors will explode as Type Ibc supernovae, many of which come from relatively low-luminosity (low-mass) regime, different from single stars or mass gainers. This is favored by inferred ejecta masses of Type Ibc supernovae \citep[e.g.,][]{Taddia2015}. Our results show that our binary model grid can be used to make comprehensive predictions for diverse supernova populations. 

At the pre-supernova stage, mass gainers do not have a companion if the binary was previously disrupted by the supernova kick, or they have a compact object as a companion. On the other hand, mass donors always have a companion, which is usually in the core hydrogen burning phase (98.2\%). Some of these systems are close enough that the outermost layers of the mass gainer will expand due to shock heating \citep{Ogata2021}. Mass gainers in $\sim 4600$ out of $38\,430$ models are expected to expand beyond its Roche lobe radius, assuming the primary becomes a neutron star without a kick. This may result in distinctive observational features associated with supernovae \citep[e.g.,][]{Chen2024}. Also, in about 20 out of $38\,430$ models, the mass gainers are supergiants when mass donors undergo core collapse. In such cases, the impact of supernova ejecta on the extended envelope of the companion can lead to notable outcomes, such as partial envelope stripping caused by shock heating \citep{Hirai2014}. Again, our model grid can offer a useful theoretical baseline for investigating such interesting phenomena and their occurrence rates. 

The models show that they can be progenitors of various types of core-collapse supernovae. RSGs will likely end up as Type IIP supernova, and stars with small hydrogen envelope mass as Type IIL/n, and hot stripped stars as Type Ibc. The models also indicate that the progenitors for each supernova type will have various properties. For example, RSGs will explode as hydrogen-rich supernova, but with various envelope masses; mass gainers have the largest envelope and mass donors have the least, which is directly related to the supernova ejecta mass. Stripped stars will explode as stripped-envelope supernova with various masses as well \citep[e.g., see][for predictions from our models]{Ercolino2025}. A recent work by \citet{Gilkis2025} presents a comprehensive overview of supernova progenitors predicted from both single star and binary channels, similar to our approach. In addition, \citet{Laplace2020} and \citet{Schneider2024} offer detailed predictions for the pre-supernova evolution and final fates of mass donors and accretors.


Here we compare our models at core carbon depletion (see \Fig{fig_preSN}) with several well-studied supernova progenitors that were detected prior to explosion. These observations, typically dating years before the explosion, provide only photometric data on the progenitors, and we will therefore only compare their positions in the HRD with respect to our model predictions.

A Type IIb supernova SN\,1993J was identified as a YSG before explosion \citep{Maund2004}. Spectroscopic and photometric analyses of the supernova revealed that only a small amount of hydrogen remained in the envelope at the time of explosion \citep{Podsiadlowski1993,Woosley1994,Houck1996}. Its position in the HRD, together with the presence of a low-mass hydrogen envelope, is consistent with our mass donor models. Notably, \citet{Maund2004} also reported evidence for a possible companion star at the explosion site, further supporting a binary evolution scenario. The companion, once thought as a BSG, is now thought to be a B-type main sequence star \citep{Fox2014}, which also support our mass donor scenario.

iPTF13bvn is the first Type\,Ib supernova whose progenitor has been detected prior to explosion \citep{Cao2013}. Its observed luminosity is inconsistent with that of a single-star Wolf–Rayet progenitor \citep{Fremling2014}, but instead suggests an origin in an interacting binary system. The best-fitting progenitor with $\logLLsun \lesssim 5$ and $ \logTeff<4.7$ \citep{Eldridge2015,Kim2015} is compatible with mass donors in our model grid. \citet{Gilkis2022} present models in which the best-fitting binary progenitor is significantly cooler, potentially retaining a hydrogen envelope of $\sim 0.01\Msun$, which is also in agreement with our models.

Many Type IIn supernovae have had progenitor detections, most of which are identified as luminous blue variables (LBVs) prior to explosion \citep{Niu2024}. These progenitors typically appear as BSGs or YSGs with $\logLLsun \gtrsim 6$, making them significantly more luminous than the brightest models in our grid. However, their known variability can be up to 1\,dex in $\logLLsun$, complicating direct comparisons. Among these, the progenitor of SN,2016jbu \citep{Kilpatrick2018} exhibits luminosities and effective temperatures that are compatible with our partially stripped mass donor models in the range $\logLLsun=4.9-5.5$. Other Type IIn supernova progenitors, like the progenitors of SN\,2009ip \citep{Smith2016} and SN\,2015bh \citep{EliasRosa2016} have luminosity variations that, that, at their faintest phases, overlap with the area occupied by our mass gainer models in the HRD (\Fig{fig_preSN}). It is important to note, however, that our models are hydrostatic and thus do not capture the variability observed in LBVs. This limitation may affect the predicted pre-explosion positions in the HRD for the most luminous mass gainers, especially the ones near the S Doradus instability strip \citep{Wolf1989,Smith2004}.

Among the RSG progenitors of Type IIP supernovae, we observe that both single stars and mass donors show an upper luminosity limit of  $\logLLsun\sim5.4$, which is still above the observational limit of $\sim 5.2$, thus not resolving the so-called "RSG Problem" \citep{Smartt2009,Smartt2015,Davies2020}. Mass gainers can reach even higher luminosities, up to $\logLLsun$ of $5.5$. \citet{Dessart2024} compared the supernova outcomes from a selected subset of our models, recomputed through to core collapse (see also \citealt{Ercolino2024}), with observations of exemplary supernovae such as SN\,1993J. These comparisons showed overall good agreement for Type Ib supernovae. However, progenitors of Type IIb and IIP supernovae in the model set tended to retain envelopes that were more massive and extended than those inferred from light curve modeling of observed Type IIP supernovae.

\subsection{Surface abundances}\label{sec_res2}

\begin{figure*}
	\centering
	\includegraphics[width=0.45\linewidth]{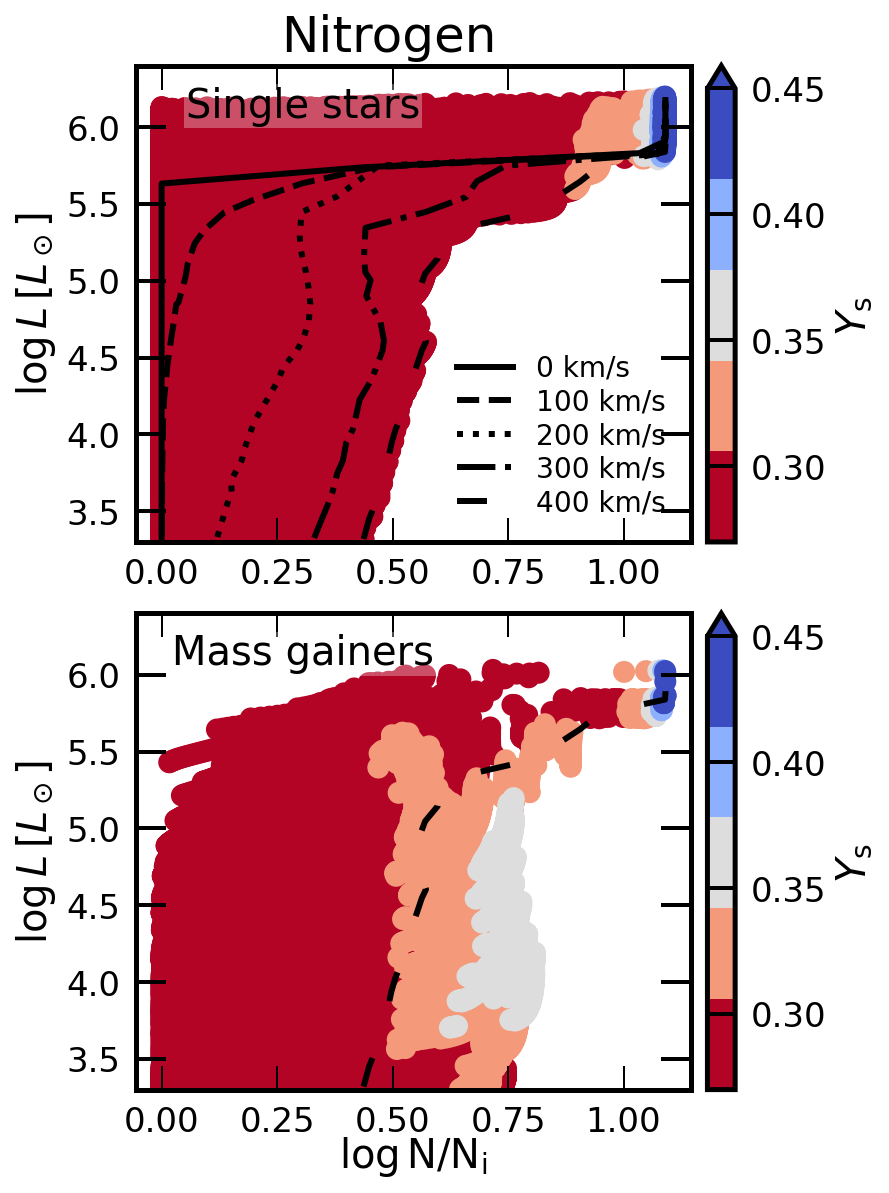}
	\includegraphics[width=0.45\linewidth]{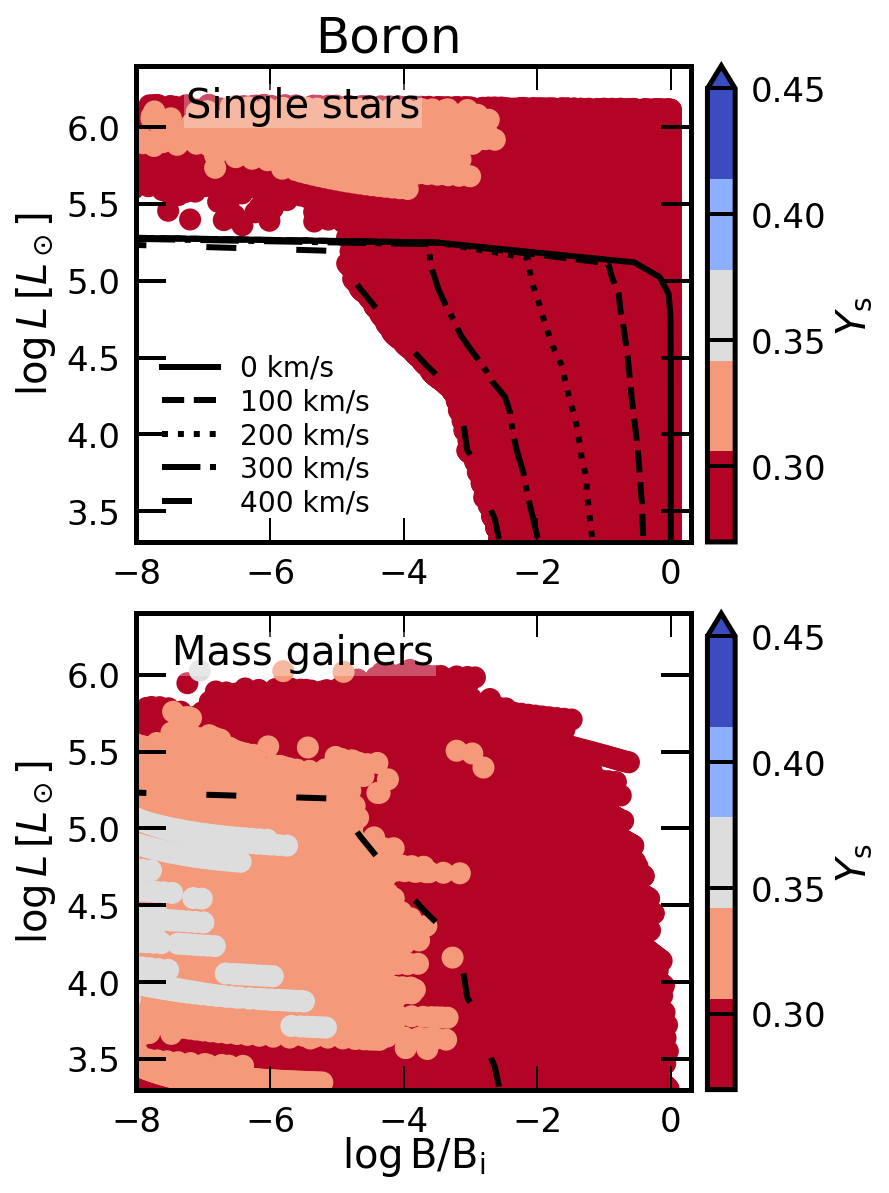}
	\caption{Surface elemental abundances of main sequence stars as a function of luminosity for nitrogen (left) and boron (right). The upper panels show single stars, while the lower panels show mass gainers. The $x$-axis represents $\log(N/N_\mathrm{i})$, where $N$ is the elemental number fraction and $N_\mathrm{i}$ is the initial value. Abundances are sampled at 0.1\,Myr intervals for all models within the main sequence band during the core hydrogen burning phase. Data points are stacked such that those with higher $\Ys$ appear above those with lower $\Ys$. For single stars, lines indicate the limits for different initial rotational velocities. For mass gainers, the line for 400$\kms$ is shown.
    }
	\label{fig_nni}
\end{figure*}

\begin{figure}
	\centering
	\includegraphics[width=0.97\linewidth]{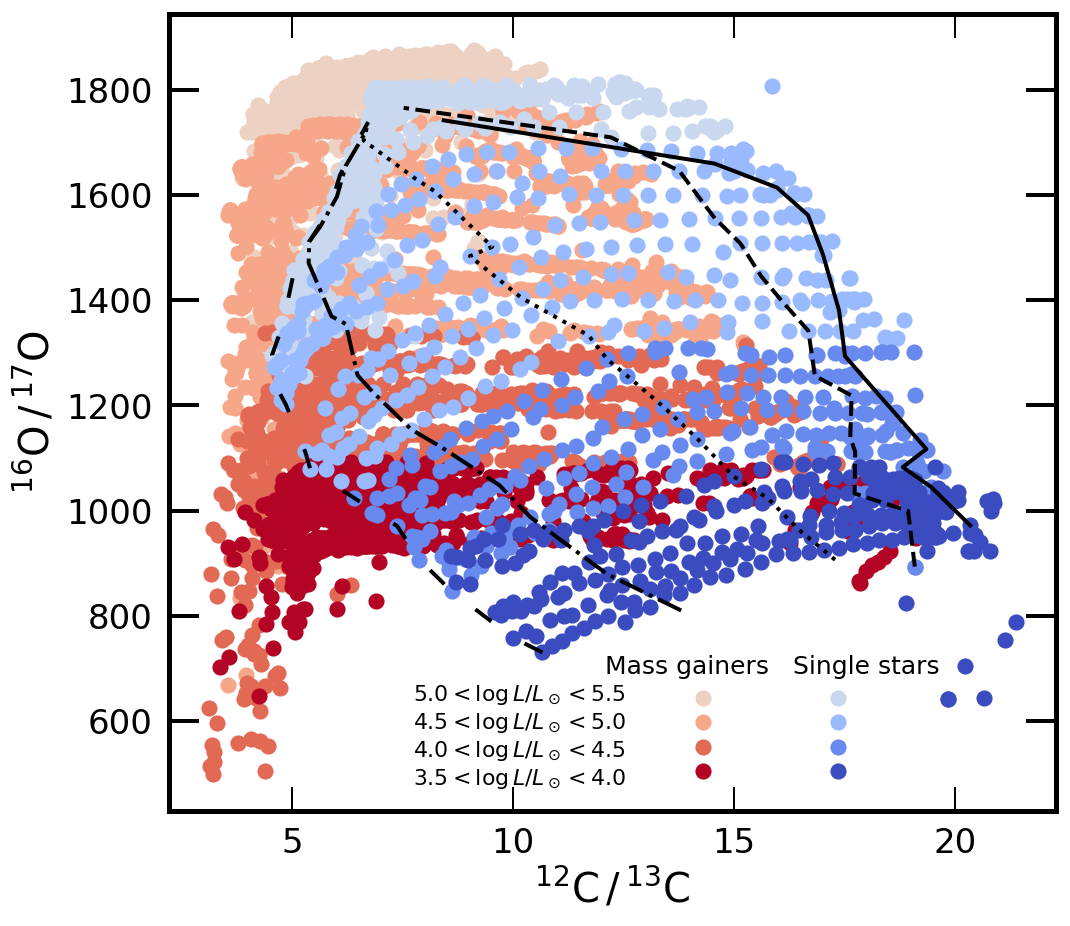}
	\caption{Predictions for $\Ciso$ and $\Oiso$ surface isotopic ratios in the red supergiant phase. Single star models and mass gainer models in different luminosity ranges are colored differently. For single star models, lines mark the limits for models with different initial rotational velocities of 0, 100, 200, 300, 400$\kms$ from right to left. The initial values of $\Ciso$ and $\Oiso$ are 90 and 2750. 
    }
	\label{fig_RSG}
\end{figure}

One of the consequences of mass transfer is a change in the surface abundances of both stellar components. For single stars, surface abundance changes may occur due to stellar wind mass loss, rotational mixing, and convective envelope dredge-up. In contrast, interacting binary stars also experience mass stripping, accretion, and accretion-induced mixing  \citep{Wellstein2001,Langer2012,Renzo2021,Wang2023}. 

The question arises whether the surface abundances of a star can be used to draw conclusions about its evolutionary history. Notably, binary interaction products are otherwise often not recognized as such, because they may have lost their companion, or they merged with it, or the companion is faint and of low mass \citep{Wellstein2001,deMink2014,Goetberg2017}. Here, we aim to simply scan through the surface abundance predictions from our comprehensive binary model grid, and compare them with corresponding predictions from single star models. A complete overview of this approach is provided in Appendix~\ref{app_abun}. In this section, we discuss three particularly interesting examples in more detail.

\subsubsection{N and B abundances in OB stars}\label{sec_nni}

\Figure{fig_nni} presents the surface nitrogen and boron abundances of OB stars of single stars and mass gainers. For stars with relatively low luminosities ($\logLLsun<5.7$), single stars exhibit stronger nitrogen enhancement for higher initial rotational velocities ($\vini$) due to stronger rotational mixing. Similar trends---either enhancement or depletion depending on the effects of nuclear burning---are observed in other elemental abundances (\Fig{fig_abun}) and isotopic ratios (\Fig{fig_iso}). In contrast, for relatively high-luminosity stars ($\logLLsun>5.7$), surface helium enhancement is significant, primarily due to strong stellar winds that strip away the outer envelope and reveal the helium-rich core.

Compared to single stars, mass gainers can exhibit stronger nitrogen enhancement at relatively low luminosities. Their surface abundances are influenced by mass accretion and accretion-induced mixing \citep[e.g.,][]{Renzo2021,Wang2023}. These stars can also experience significant spin-up due to the accretion of angular momentum, which may induce rotational mixing (see \Fig{fig_al_vrot}), and may be observed as Be stars. On the other hand we notice that a nitrogen enhancement exceeding that found in single stars occurs mostly in models with significant surface helium enrichment, where it can be higher than that in single stars with very high initial rotation velocities as 400$\kms$. If the initial rotational velocities of OB stars follow the rotational velocity distribution proposed by \citet{Dufton2013}, at most $<9\%$ of stars are born with $\vini>400\kms$, while many of th fast rotators may be binary products. In other words, only a small fraction of stars can reach nitrogen enhancement levels of $\logNNi\gtrsim0.5$ (as seen in mass gainers) through rotational mixing alone. Additionally, at relatively low luminosities, single stars do not show helium enhancement, while mass gainers do due to the accreting helium-rich material. These mass gainers exhibit interesting CNO abundance patterns (Jin et al. In prep.). At higher luminosities ($\logLLsun>5.7$), mass gainers show similar behavior as single stars, as their surface abundances is primarily determined by strong wind.

One other sensitive tracer of rotational mixing, stellar winds, and binary effects is the surface boron abundance \citep[e.g.,][]{Fliegner1996,Venn2002,Frischknecht2010,Jin2024}. Stars at lower luminosity, say $\logLLsun \sim 4$, show a factor of $\sim1000$ difference between non-rotating and fastest rotating models in surface boron abundance, while nitrogen shows a difference of a factor of three, showcasing the boron's sensitivity to rotational mixing. The effects of stellar winds on surface boron become dominant starting at lower luminosity ($\logLLsun\sim5.1$) than that of nitrogen ($\logLLsun\sim5.7$), arises from the fact that boron depletion occurs closer to the surface than nitrogen enhancement. Mass gainers exhibit stronger boron depletion than single stars. Especially, the nitrogen-rich and helium-rich ones exhibit boron depletion factor of $\lesssim10\,000$ so they cannot be detected \citep[see][]{Kaufer2010,Jin2024} in the spectra. 

The surface abundances of boron and nitrogen have been shown to be powerful diagnostic of identifying products of binary interaction, effectively disentangling them from stars that evolved in isolation. This allowed the effects of single star evolution---such as rotational mixing---to be isolated, thereby enabling getting constraints on them \citep{Jin2024}. 

\subsubsection{C and O isotope ratios in RSGs}

In cool stars, isotope ratios can be measured. Similar to elemental abundances, isotopic abundances also change following mass transfer. Carbon and oxygen isotopes are sensitive to the CNO-cycle, and while the numbers of $\Ca$ and $\Oa$ nuclei can decrease due to the CNO-cycle, those of $\Cb$ and $\Ob$ can increase in certain regions under appropriate temperature conditions. 

\Figure{fig_RSG} shows the ratios of $\Ciso$ and $\Oiso$ for our RSG models. In table\,B.1 in \citet{Asplund2021}, from which we adopt the initial abundances, the carbon isotopic ratio is taken from \citet{Meija2016}, and the oxygen isotopic ratios are adopted from \citet{McKeegan2011}. In single stars, the faster the star rotates initially, the stronger the changes in the isotopic ratio
(i.e., lower $\Ciso$ and $\Oiso$) due to stronger rotational mixing. Across different luminosities, the less luminous stars show lower $\Oiso$. We find that this is because $\Ob$ is more enhanced in less massive stars at the bottom of the envelope during the main sequence phase. Mass gainers exhibit lower $\Ciso$ compared to single stars as their envelope is contaminated with more CNO-processed material. Thus, if an observed RSG has a very low $\Ciso$ value, it might indicate that this star has accreted mass from a companion star previously. Note that the thermonuclear reaction rate of $\Ob (p,\gamma)$ is also quite uncertain \citep[e.g.,][]{Fox2005}, which is known to affect the surface oxygen isotopic ratios \citep[e.g.,][]{Lebzelter2015}.  

\subsubsection{Supernova progenitors}

At the pre-supernova stage, the surface abundances of single stars are also different compared to post-mass transfer stars. The bottom panels of \Fig{fig_preSN} compare $\logNC$ between single stars and post-mass transfer stars. For the nitrogen-to-carbon ratio ($\logNC$), its initial value is $\sim -0.6$ the CNO-equilibrium value is $\sim 2.0$. 

Single stars in the hotter branch ($\logTeff\gtrsim 5.0$) show lower $\logNC$ values than the initial value as they revealed their bare helium cores down to the region where carbon is enhanced and nitrogen is reduced. In the cooler branch ($\logTeff \lesssim 5.0$), stars with relatively high luminosities ($\logLLsun \gtrsim 5.5$) exhibit $\logNC$ values consistent with CNO-equilibrium, while those with relatively low luminosities ($\logLLsun \lesssim 5.5$) only show moderate nitrogen enhancement. These reflect different physical processes: higher luminosity stars reveal deep layers due to strong winds, while lower luminosity stars are affected by previous dredge-up event. 

Mass gainers on the cooler side generally exhibit higher $\logNC$ values, resulting from the accretion of CNO-processed material and accretion-induced mixing. In contrast, mass donors show a range of $\logNC$ values, depending on the extent of envelope stripping they have undergone during the mass transfer event. On the hotter branch, both mass gainers and donors exhibit $\logNC$ values similar to single stars. 

If $\logNC$ can be inferred from supernova ejecta---reflecting the surface composition at the pre-supernova stage---it may offer valuable insights into the progenitor's evolutionary history, particularly in distinguishing between stars that evolved as single or have experienced binary mass transfer events.

\section{Discussion}\label{sec_dis}

\subsection{Previous surface abundance predictions for post-mass transfer stars}\label{sec_comp}

Massive binary evolution models, in which both stellar components and the orbit are followed in detail, have been computed by various groups over the last few decades. Some of them also provide predictions for the surface abundances of post-mass transfer stars. 
One of the earliest such works is by \citet{Wellstein1999,Wellstein2001}, and they found surface helium and nitrogen enhancements of which factors are comparable to our predictions (\Fig{fig_abun}). Notably, all of their mass gainers are BSGs during core helium burning—likely a result of their assumption of fully conservative mass transfer, which differs from our approach. \citet{Petrovic2005,Song2018mt} investigated surface helium and nitrogen abundances of both mass gainers and donors, which are also comparable with our models. Works by \citet{Langer2008, Langer2010, Langer2012} showed the depletion of light elements like boron and beryllium in mass gainers, highlighting the sensitivity of light elements to mass transfer. 

\citet{Renzo2021} presents a detailed investigation of a mass gainer model, which could reproduce the properties of $\zeta$ Ophiuchi. This star is a fast-rotating, runaway star with helium and nitrogen enhancement at the surface, which has been proposed to be a mass gainer. They showed that accretion of CNO-processed material and ensuing accretion-induced mixing (thermohaline mixing and rotational mixing) alters the internal chemical profiles and surface abundances of mass gainers, and could reproduce the observed surface abundances of the star. Fast rotation and the runaway characteristic of the star also aligns well with the binary supernova scenario.

The Bonn stellar group has constructed detailed binary evolution model grids for three different metallicity environments: SMC \citep{Wang2020}, LMC \citep{Langer2020, Pauli2022}, and the Galaxy (this work)\footnote{The input files and the whole model data from Bonn SMC and LMC binary grids are also available on the same platform as Bonn GAL.}. They share similar physics assumptions, but a key difference is the nuclear network: the Bonn GAL models use a much more detailed network for hydrogen burning, while the Bonn LMC and SMC models use simpler network. Based on the grids, \citet{Pauli2022} provides surface helium mass fraction for binary stripped Wolf-Rayet stars \citet{Xu2025} for mass gainers with compact companions. \citet{Sen2022, Sen2023} investigated surface abundances of mass gainers and donors in semi-detached systems (i.e., systems undergoing slow Case A mass transfer). Mass donors reveal inner layers sequentially as mass transfer goes on, and the surface nitrogen eventually reaches near the CNO-equilibrium value. Their mass gainers also show helium and nitrogen enhancement at the surface (see \Fig{fig_abun} for our case).

\subsection{Limitations of our predictions}\label{sec_lim}

As our binary grid uses a single set of input physics, we do not have models across possible physics assumptions, so the models' results should be used with caution. For example, the rotation-limited accretion adopted in our models results in a low accretion efficiency for wide binaries. Observational evidence from post-mass-transfer binaries suggests higher accretion efficiencies than those in our models for relatively low-mass binary systems \citep[e.g.,][in which they find that accretion efficiency can be up to or even larger than 50\%]{Vinciguerra2020,Xu2025,Schuermann2025}. Based on our binary models, \citet{Carlos2025} showed the potential of using surface abundances to constrain the accretion efficiency. Since higher accretion efficiency leads to stronger changes in surface abundance of the mass gainers, surface helium is more enhanced in accretors under a conservative mass transfer scenario. With a more conservative mass transfer, we expect that mass gainers, as shown in Figs.~\ref{fig_abun} and \ref{fig_iso}, would exhibit larger differences when compared to single stars.

The initial rotation rate, set to 20\% of the critical rotation in our binary models, is lower than the main peak value of around 60\% \citep{Dufton2013}. The rotation of the two stars after mass transfer is determined by the interaction, so the choice of initial rotation is less critical if we just care about post-mass transfer rotation rates. However, the effects of rotation before mass transfer, particularly rotational mixing, depend on this initial choice. In our case, the effects of rotational mixing might be underestimated, so the surface abundances shown in \Fig{fig_abun} should be viewed as a conservative estimate. Nevertheless, the findings from \citet{Jin2024} support a rotational mixing efficiency that is 50\% lower than currently adopted, potentially limiting this underestimation.

The location of core helium burning donor stars in the HRD is relatively well-constrained, as it mainly depends on how much of their envelope has been stripped. In contrast, that of mass gainers depends on several processes that are less well-understood. For instance, the efficiency of mass accretion will influence the post-mass transfer luminosities. Additionally, assumptions on semiconvection affect the core growth of the secondary component upon mass transfer, which in turn influences its further evolution. This is particularly important in determining whether the secondary undergoes core helium burning as a BSG \citep{Braun1995,Saio2013,Georgy2014,Schootemeijer2019,Kaiser2020}. 

Another binary interaction process for which we do not provide predictions, but which can significantly alter surface abundances, is stellar merger. Even though we do not compute merger models, our results may provide useful insights into how merger products would behave for two reasons. First, the envelope of merger products is likely a mixture of the envelope materials of both stars and the core material of either the primary or secondary \citep{Schneider2019,Menon2024}. As a result, a significant amount of helium-rich CNO-equilibrium matter will be incorporated into the envelope of the merger products. Therefore, mergers can show strong surface helium enhancement. Post-main sequence merger models from \citet{Menon2024} illustrates this. Second, mergers can be BSGs during core helium burning, similar to that of our mass gainers due to the large envelope-to-core mass ratio \citep{Justham2014,Schneider2024}. Distinguishing between merger products and mass gainers based on orbital motions might be challenging, as both are likely single stars at this stage; however, differences in surface abundances may offer a diagnostic.

The abundances of the light elements (lithium, beryllium, boron) can be a good tracer to identify binary interaction products as well. These elements are destroyed at relatively low temperatures (roughly $3-7$ MK), near the stellar surface. Thus, their surface abundances are highly sensitive to internal mixing processes as well as binary interactions, allowing for a critical test of these processes. \citet{Jin2024} showed that many of the very slow rotators (with rotational velocity of $< 50 \, \rm km/s$) in their sample could be merger products as they exhibit significant surface boron depletion. This highlights the importance of the surface abundances of hydrogen-burning products—not only helium and CNO elements, but also the light elements—as key indicators for unraveling the evolutionary histories of stars.

\section{Conclusions}\label{sec_con}

We have presented a new grid of binary evolution models which incorporate detailed binary interaction physics and an extensive nuclear network for hydrogen burning. Our models provide insights into the evolution of post-mass transfer stars---their evolution in the HRD and surface abundances, which can serve as tracers of their binary mass transfer history.

We find that post-mass transfer stars show a wide spread in the HRD during the core helium burning phase. Mass gainers undergo core helium burning as BSGs if they accrete substantial amounts of mass. Mass donors in most cases become stripped helium stars or Wolf-Rayet stars, but partially-stripped ones show a wide range of effective temperatures during the core helium burning phase. 

Our post-mass transfer models result in progenitors for various types of supernova, ranging from Type IIP to Type Ibc. Supernova progenitors that experienced mass transfer event show distinctive properties (e.g., remaining envelope masses, chemical composition) compared to those that evolved in isolation. The properties of our partially-stripped mass donors at the pre-supernova stage match those inferred for the progenitor of SN\,1993J, highlighting the predictive power of our models. Overall, our binary model grid offers a great theoretical baseline for exploring supernova populations and related phenomena such as interacting supernovae.

Due to the accretion of nuclear processed matter and the consequent mixing, mass gainers show different surface abundances compared to single stars. In particular, they show stronger helium and nitrogen enhancement, along with stronger boron depletion than single stars. Supergiants that evolved in isolation or accreted mass in binaries also exhibit distinctive isotopic ratios of carbon and oxygen. Furthermore, supernova progenitors that have undergone mass transfer exhibit different CN abundances, reflecting their binary interaction history. Mass donors exhibit surface signatures of processed material at the surface due to binary mass stripping, with the extent of these signatures depending on the degree of mass stripping, during core helium burning and at core collapse.

Our models provide comprehensive predictions for elemental abundances and isotopic ratios across the HRD, offering a great baseline for interpreting the surface properties of stars. Future comparisons between the surface properties of these models and stars can yield valuable insights into their evolution histories. Conversely, such observational data can be used to test and refine the physics assumptions underlying in models. In a forthcoming paper, we will present a detailed study on the surface CNO abundances, compare our predictions with observations, and demonstrate how surface abundance patterns can be used to constrain key uncertainties in stellar and binary physics.

\begin{acknowledgements}
We thank Alexander Heger, Pablo Marchant, Chen Wang, Sergio Simon-Diaz, Lee R. Patrick, Thibault Lechien, 
Götz Gräfener for their help. HJ received financial support for this research from the International Max Planck Research School (IMPRS) for Astronomy and Astrophysics at the Universities of Bonn and Cologne. The authors gratefully acknowledge the granted access to the Bonna cluster hosted by the University of Bonn.
\end{acknowledgements}

\bibliographystyle{aa}
\bibliography{Reference_list}

\begin{appendix}

\section{Overview our binary evolution model grid}\label{app_Pq}

\begin{figure}
	\centering
	\includegraphics[width=\linewidth]{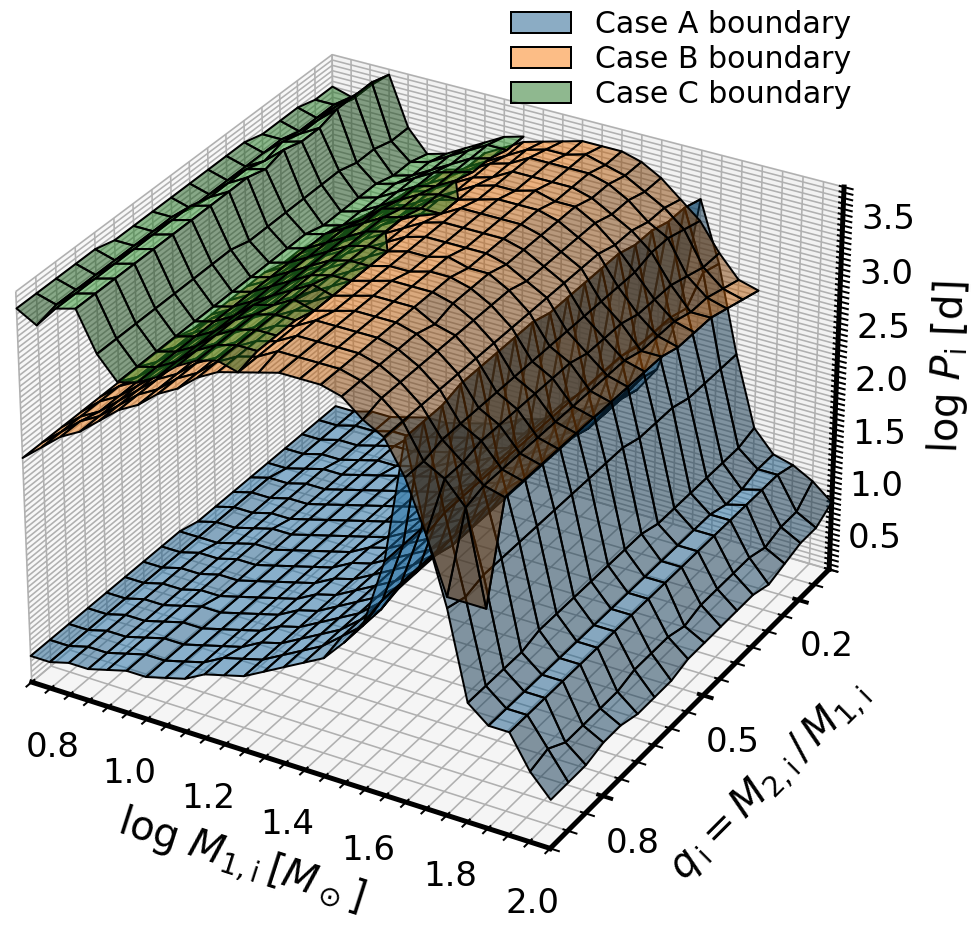}
	\caption{The boundaries of Case A (blue), Case B (orange), and Case C (green) across the whole initial binary parameter space of our grid. Binary systems below each boundary undergo the corresponding case(s) of mass transfer event.}
	\label{fig_3D}
\end{figure}

\begin{figure*}
	\centering
	\includegraphics[width=0.95\linewidth]{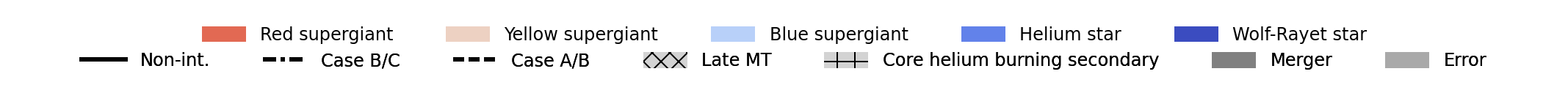}
	\includegraphics[width=0.24\linewidth]{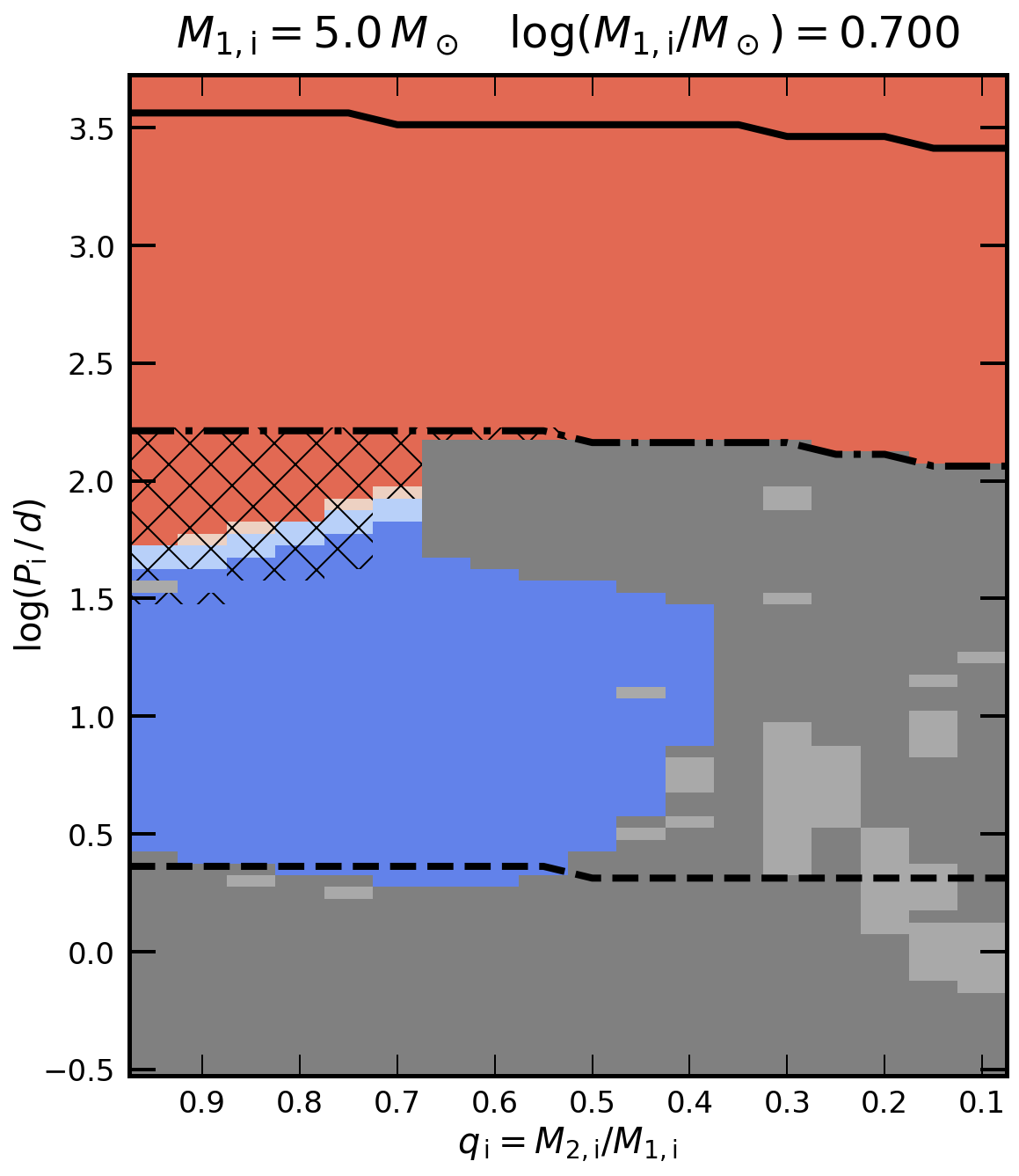}
	\includegraphics[width=0.24\linewidth]{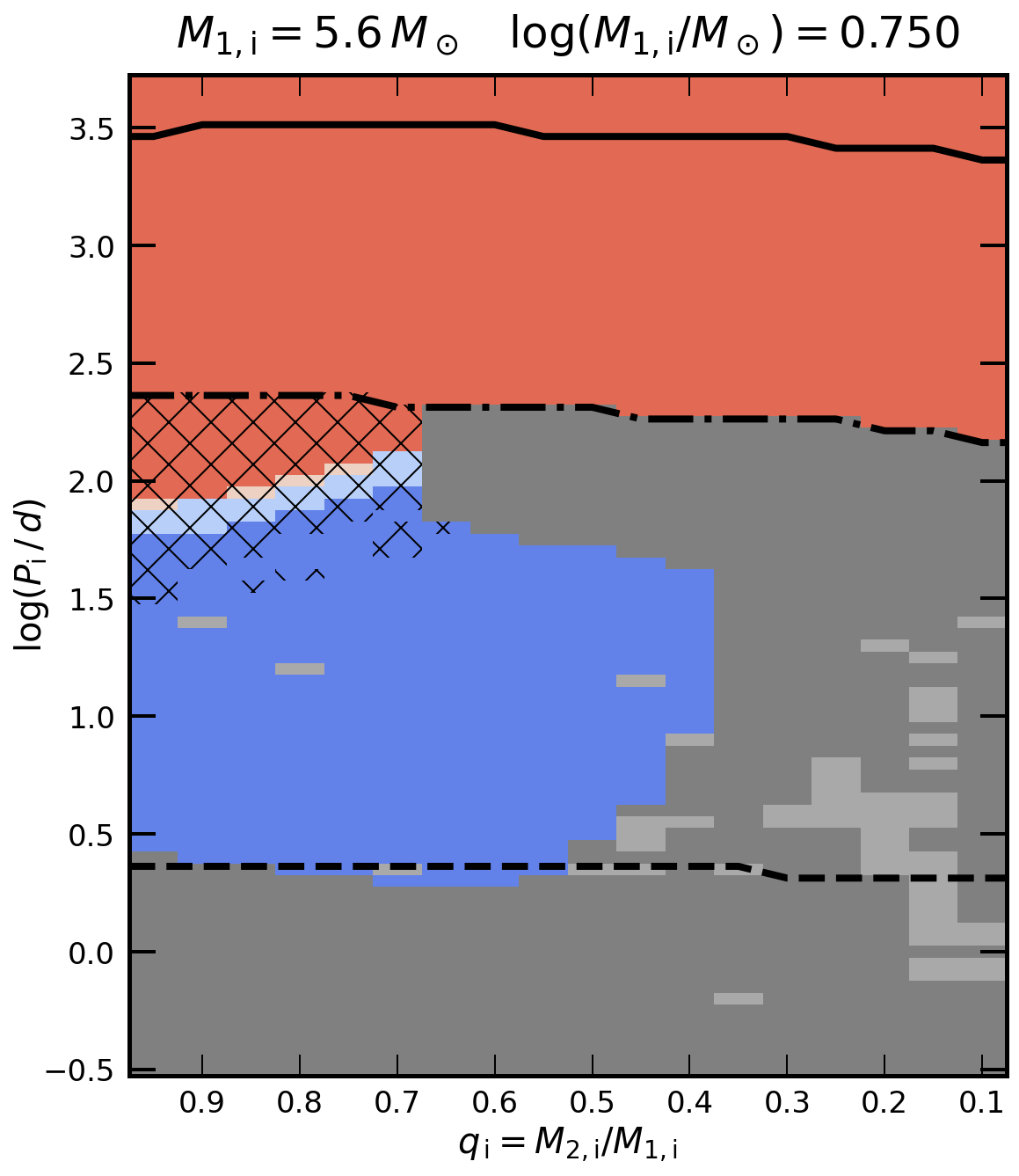}
	\includegraphics[width=0.24\linewidth]{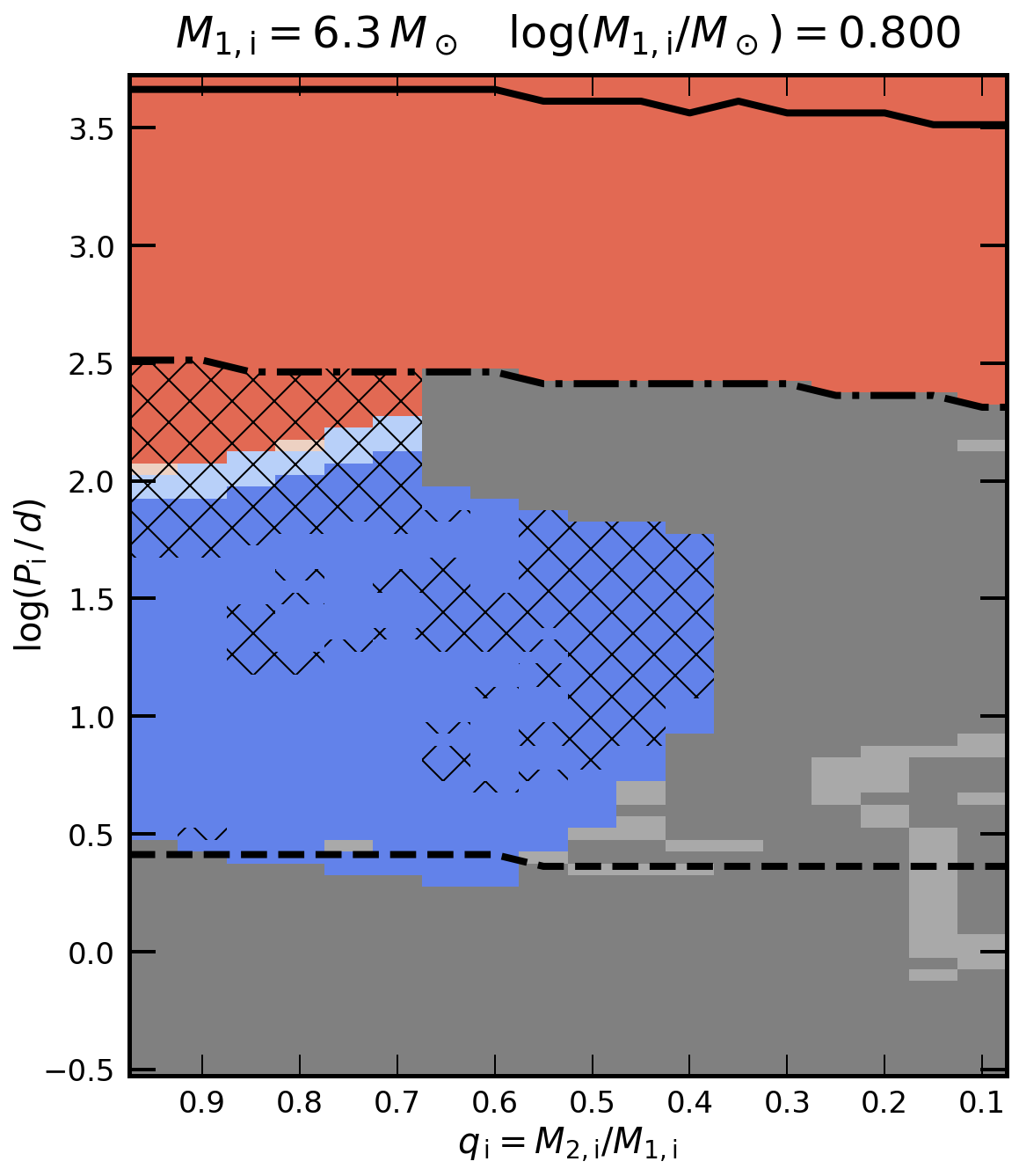}
	\includegraphics[width=0.24\linewidth]{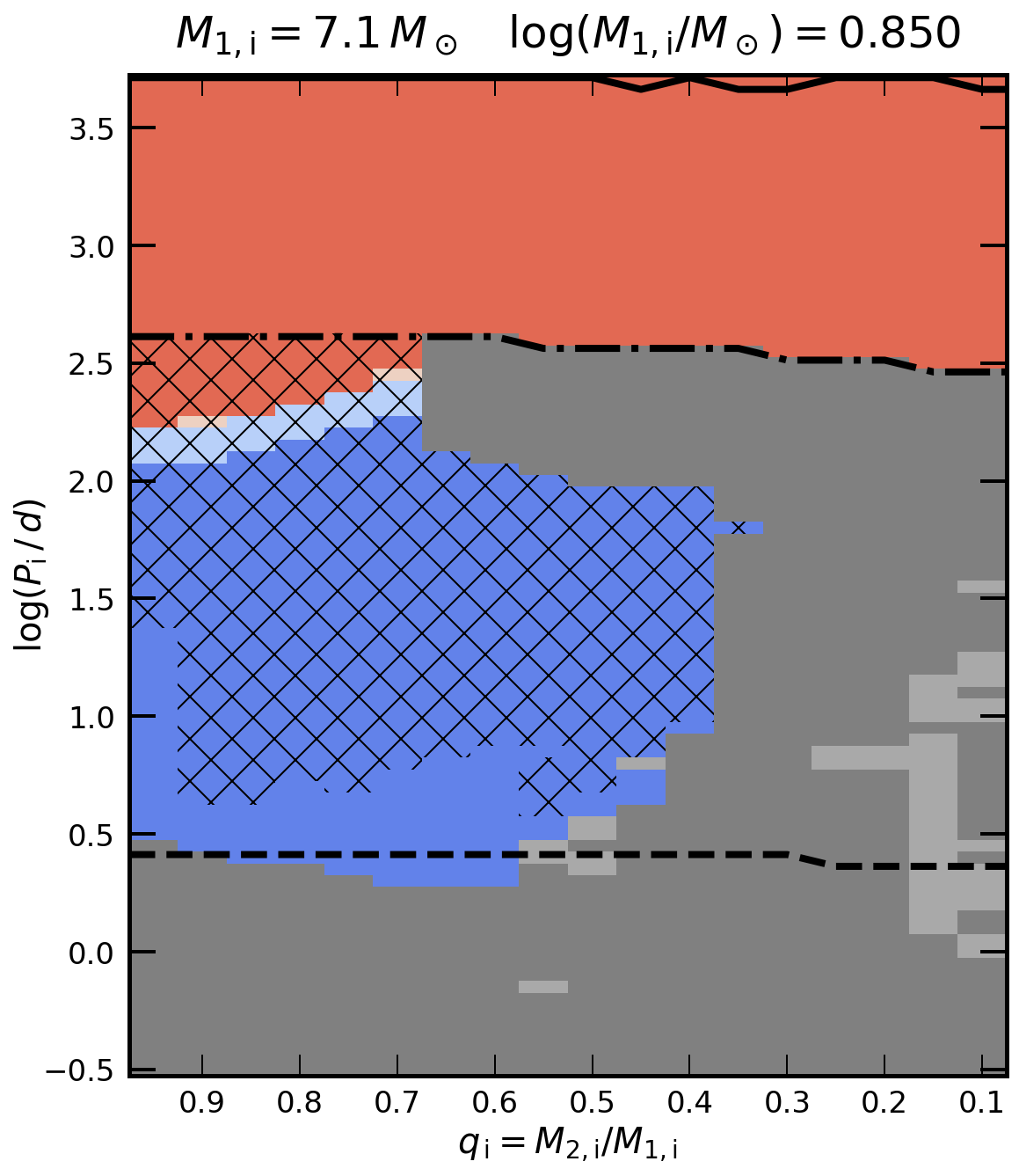}
	\includegraphics[width=0.24\linewidth]{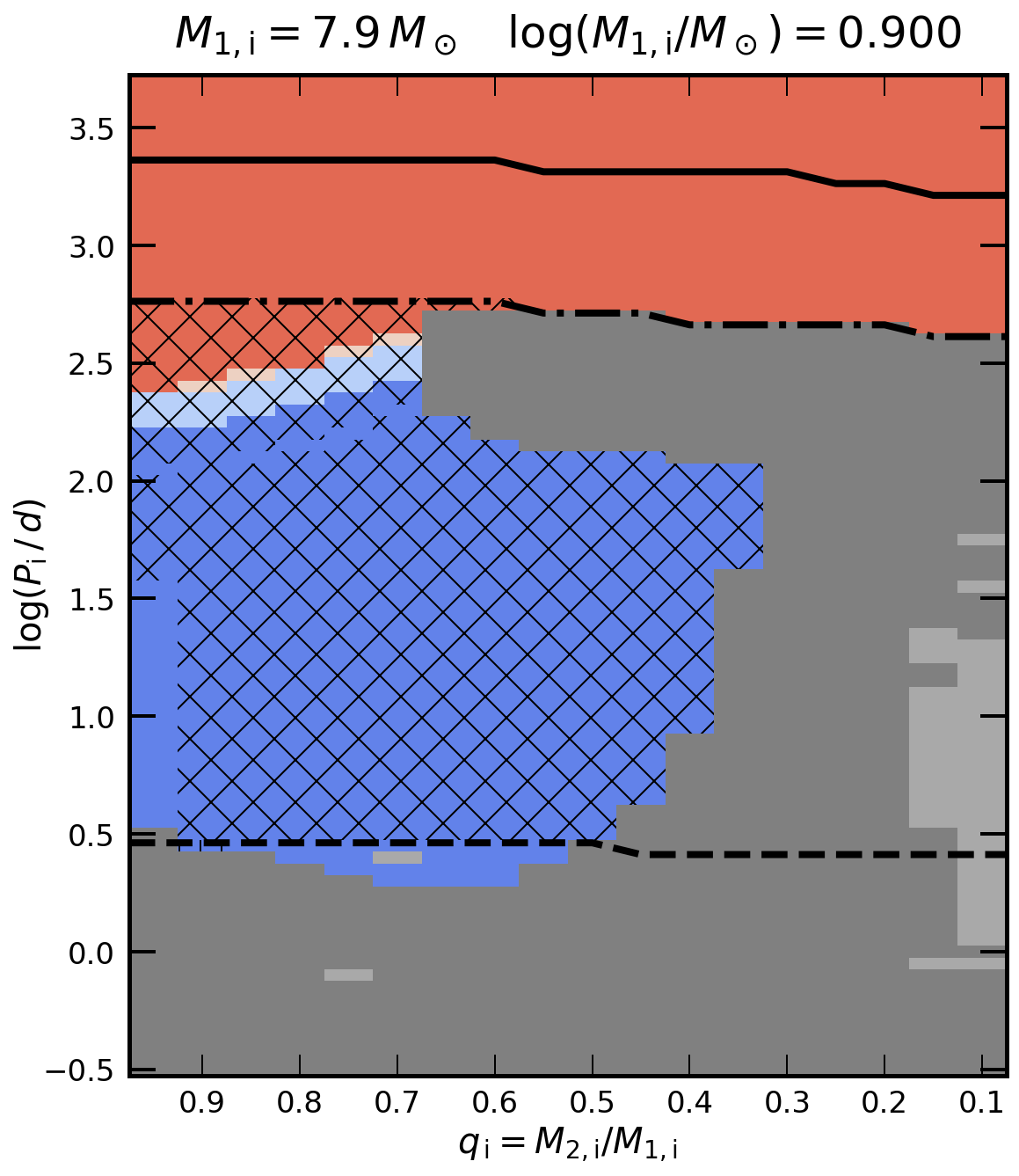}
	\includegraphics[width=0.24\linewidth]{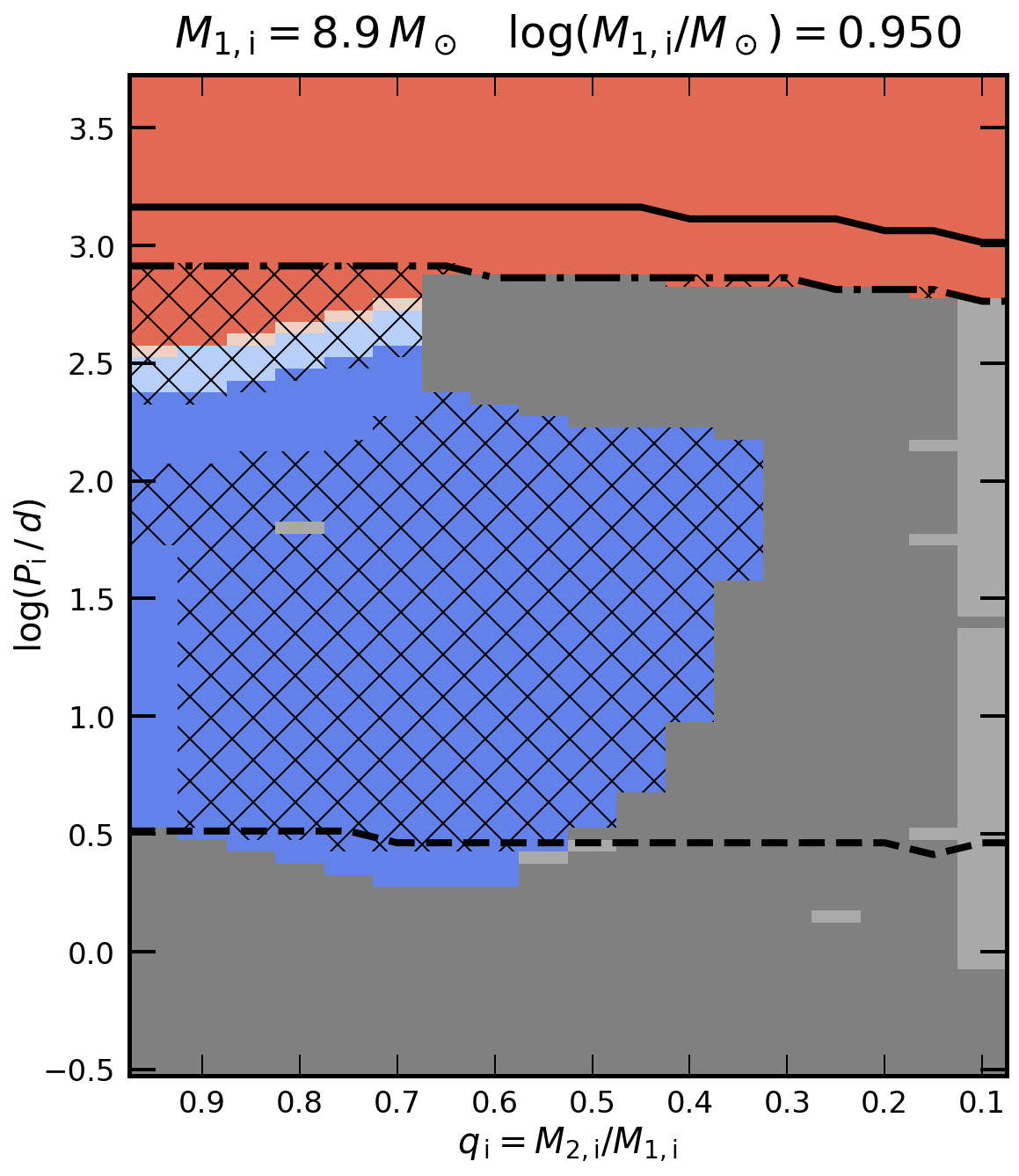}
	\includegraphics[width=0.24\linewidth]{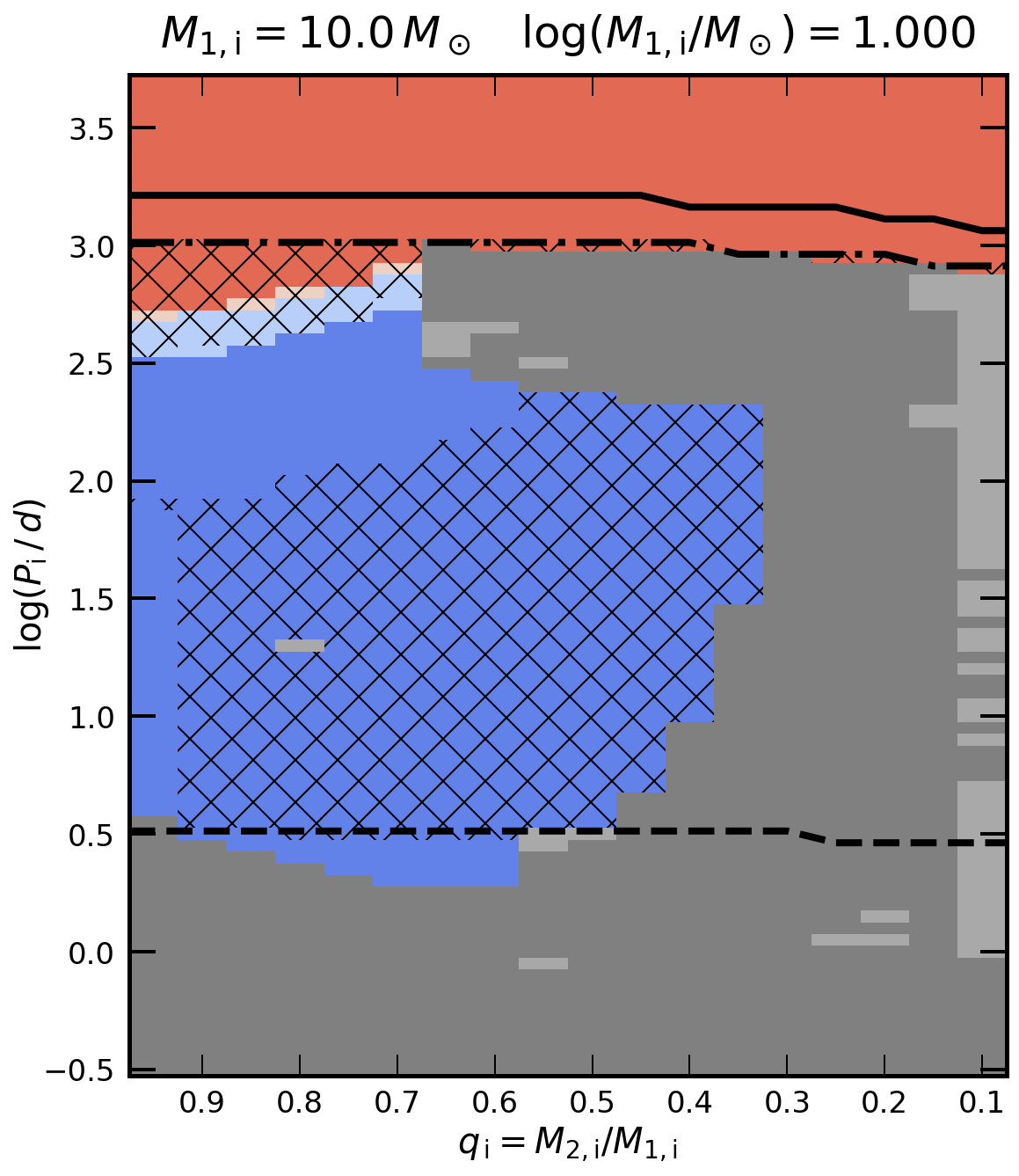}
	\includegraphics[width=0.24\linewidth]{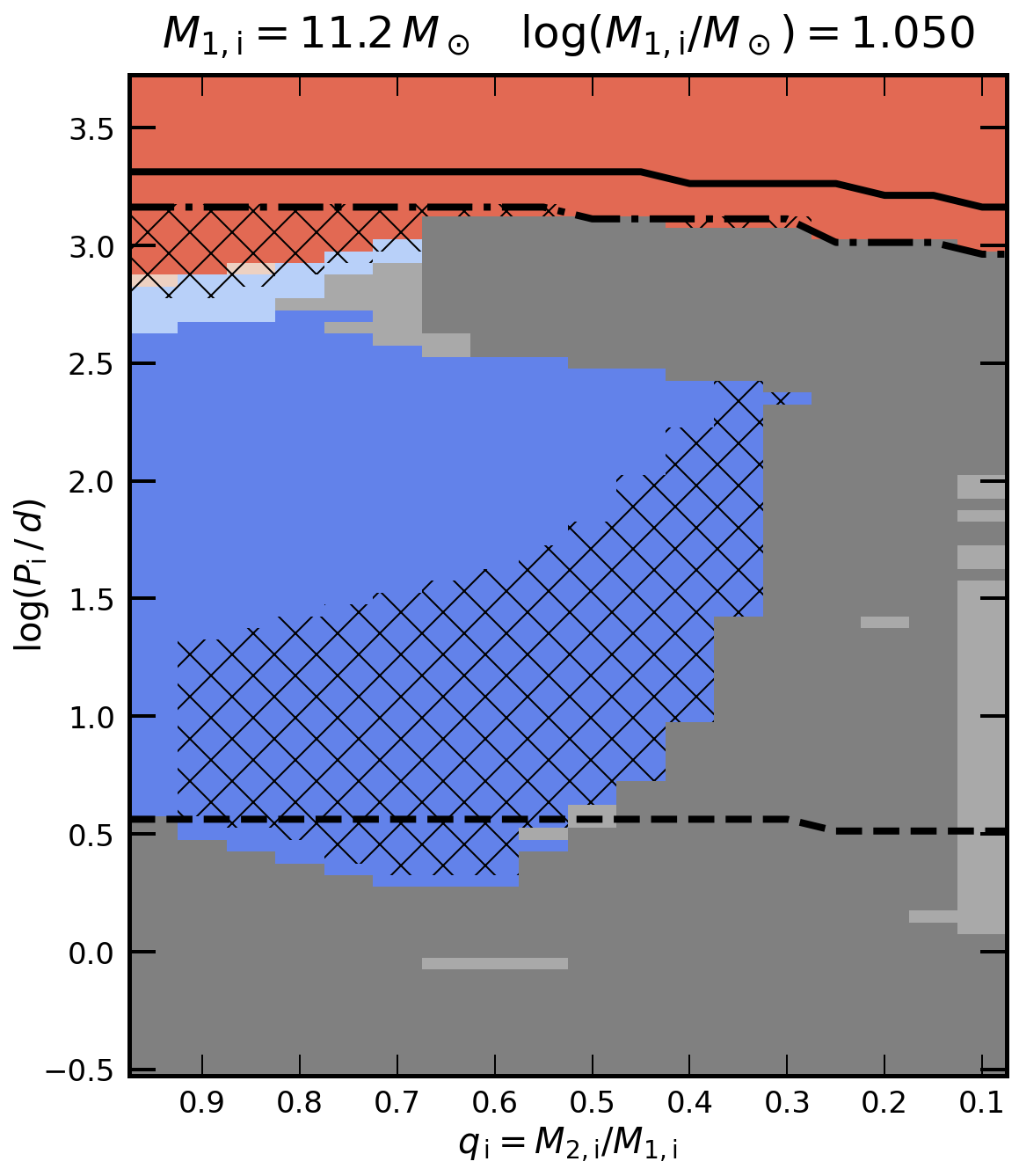}
	\includegraphics[width=0.24\linewidth]{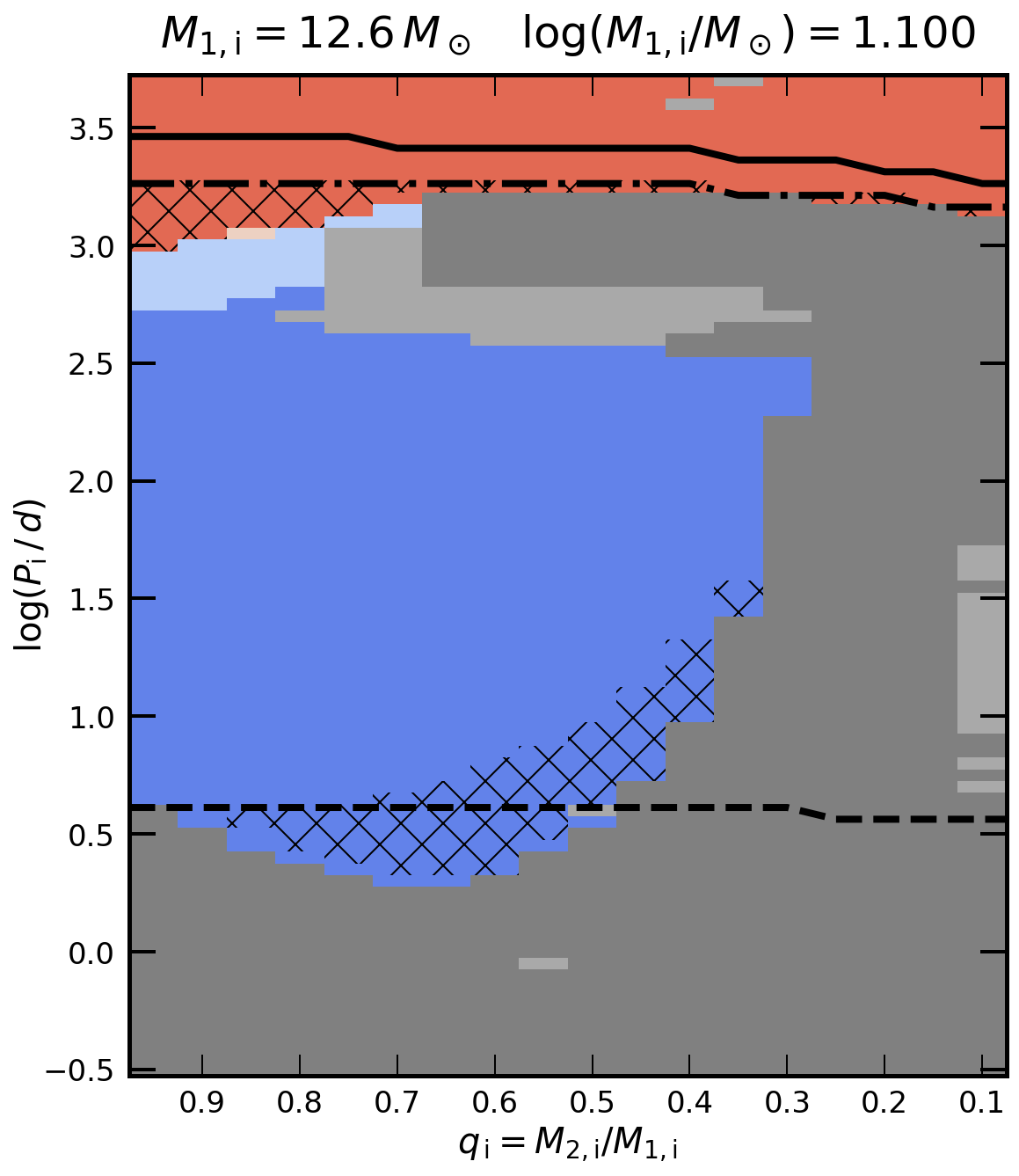}
	\includegraphics[width=0.24\linewidth]{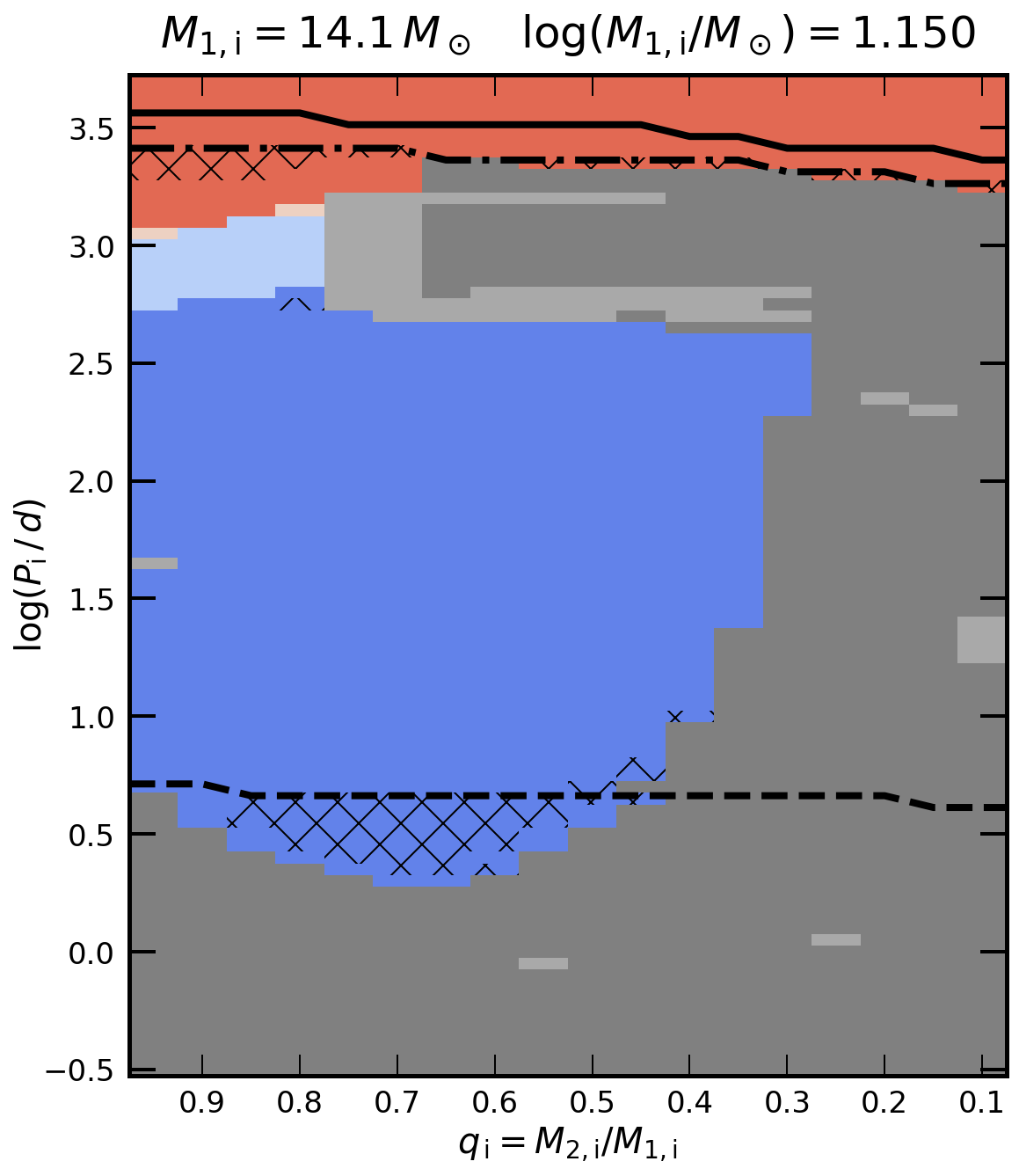}
	\includegraphics[width=0.24\linewidth]{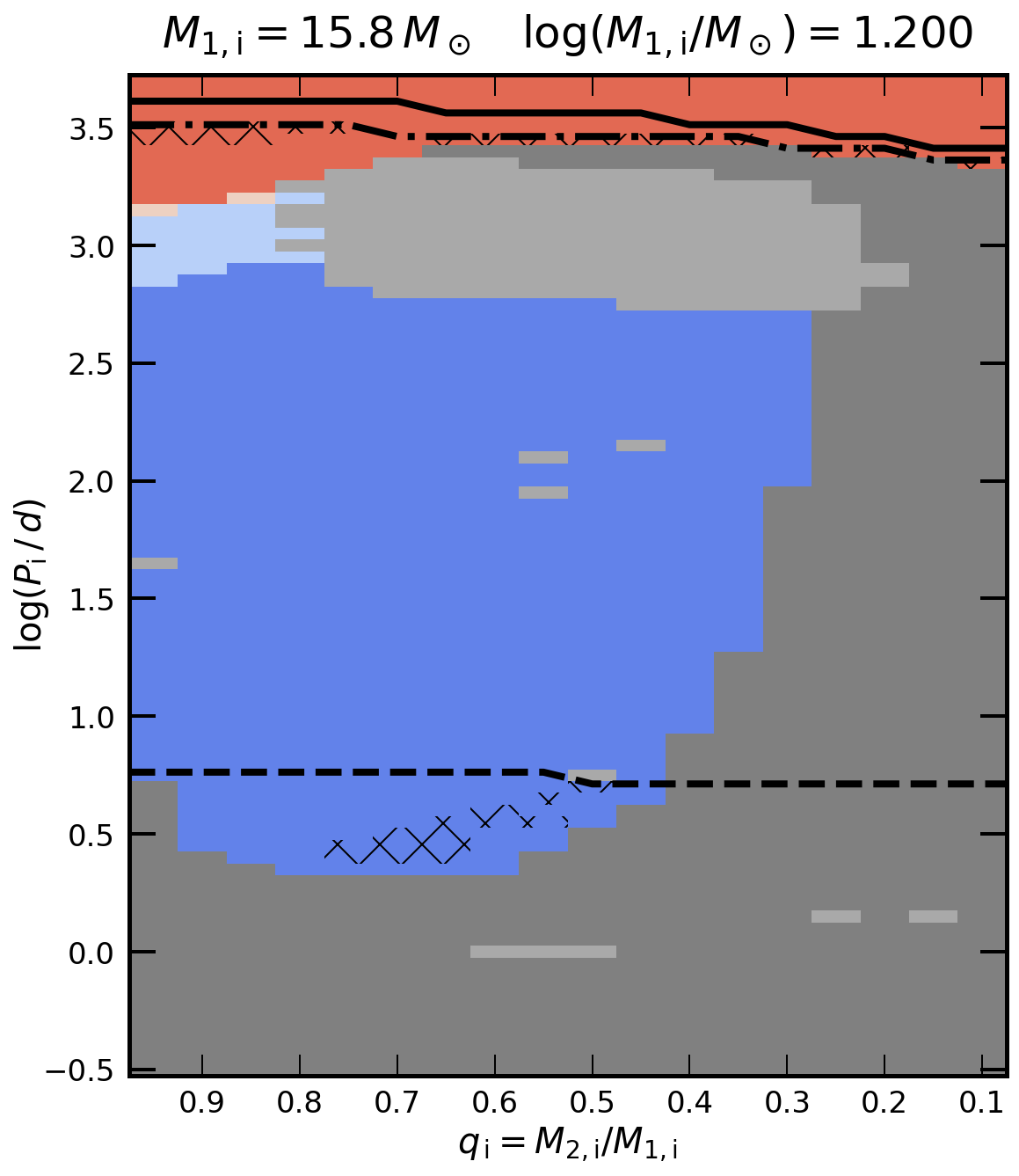}
	\includegraphics[width=0.24\linewidth]{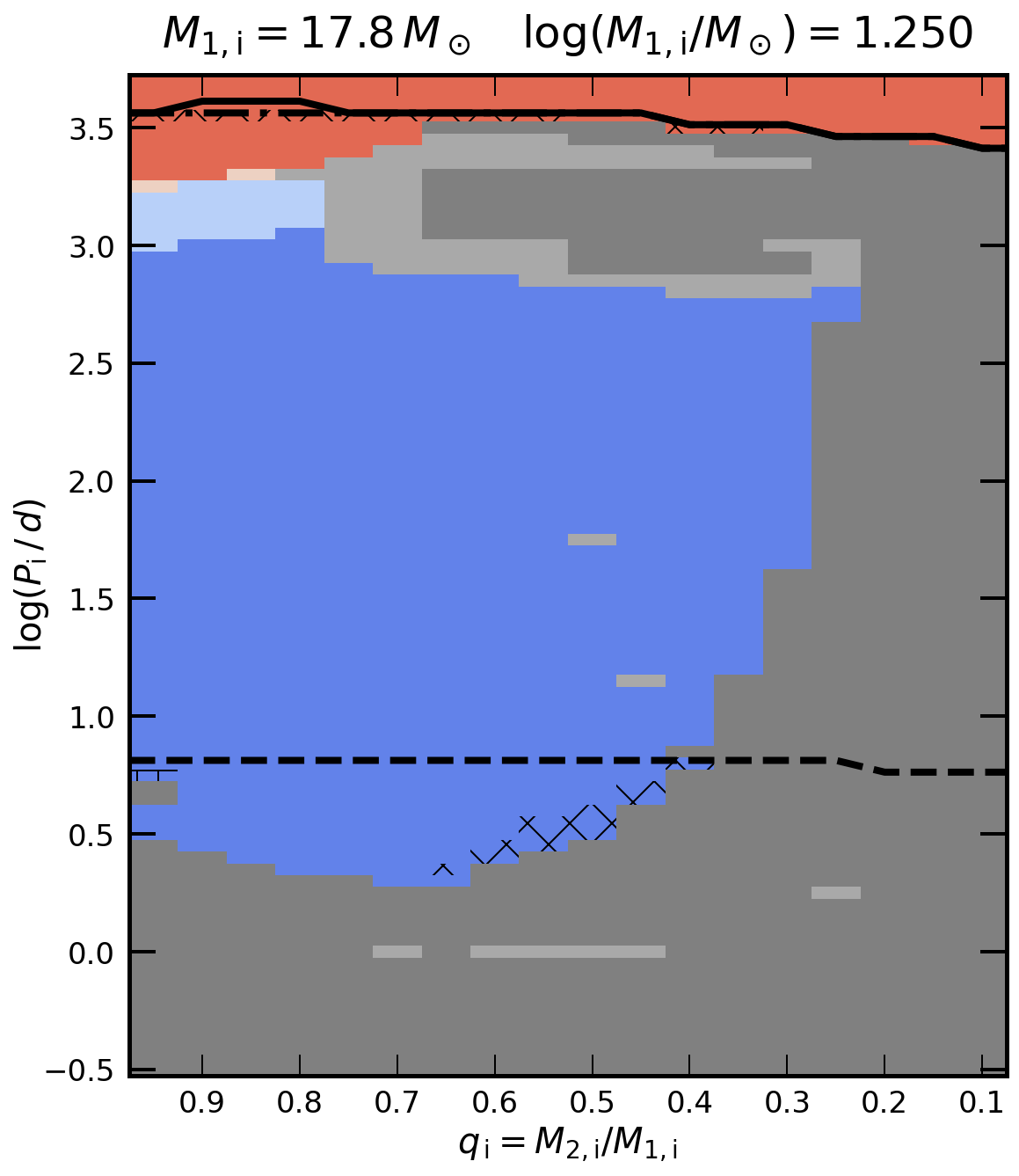}
	\includegraphics[width=0.24\linewidth]{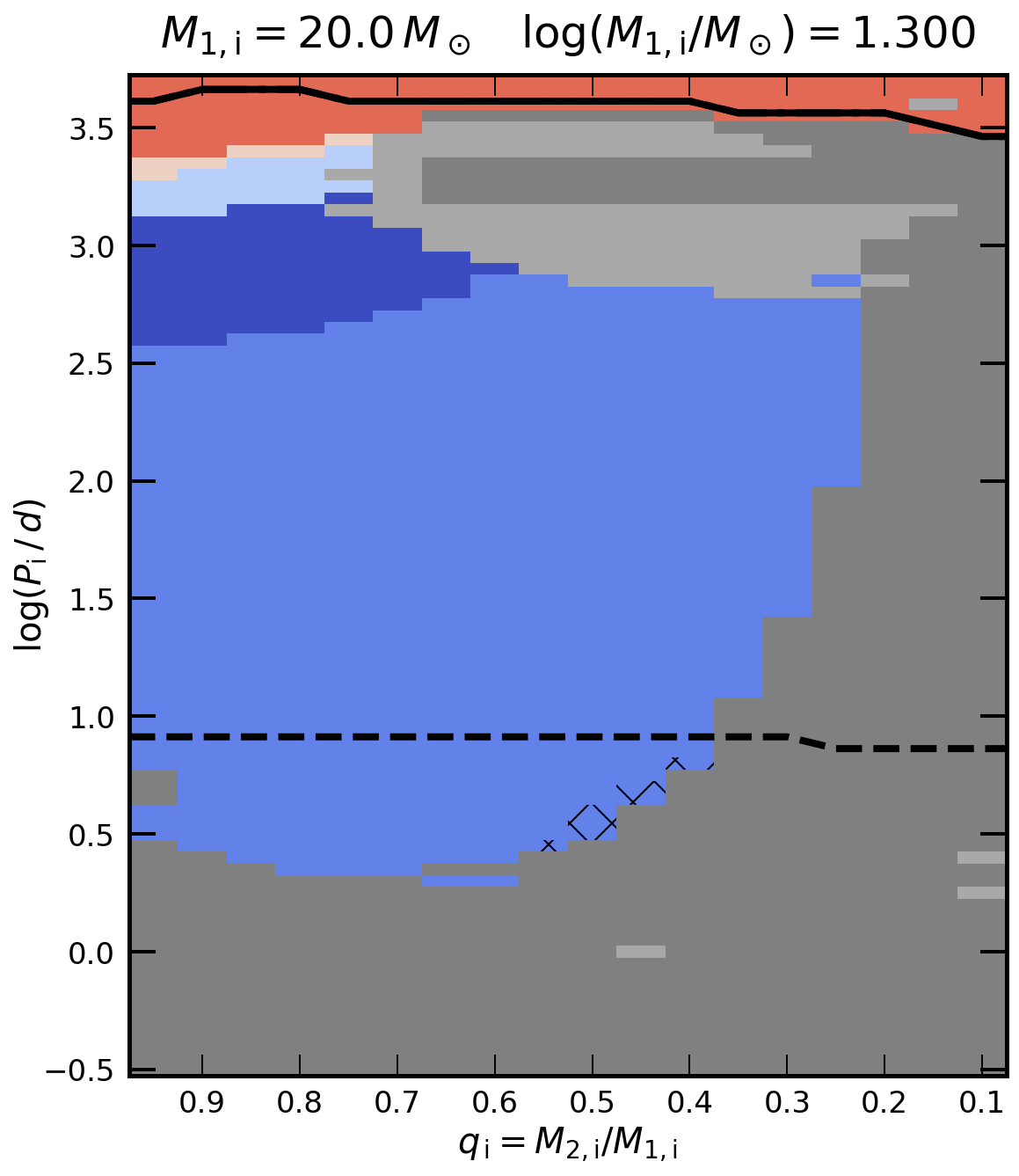}
	\includegraphics[width=0.24\linewidth]{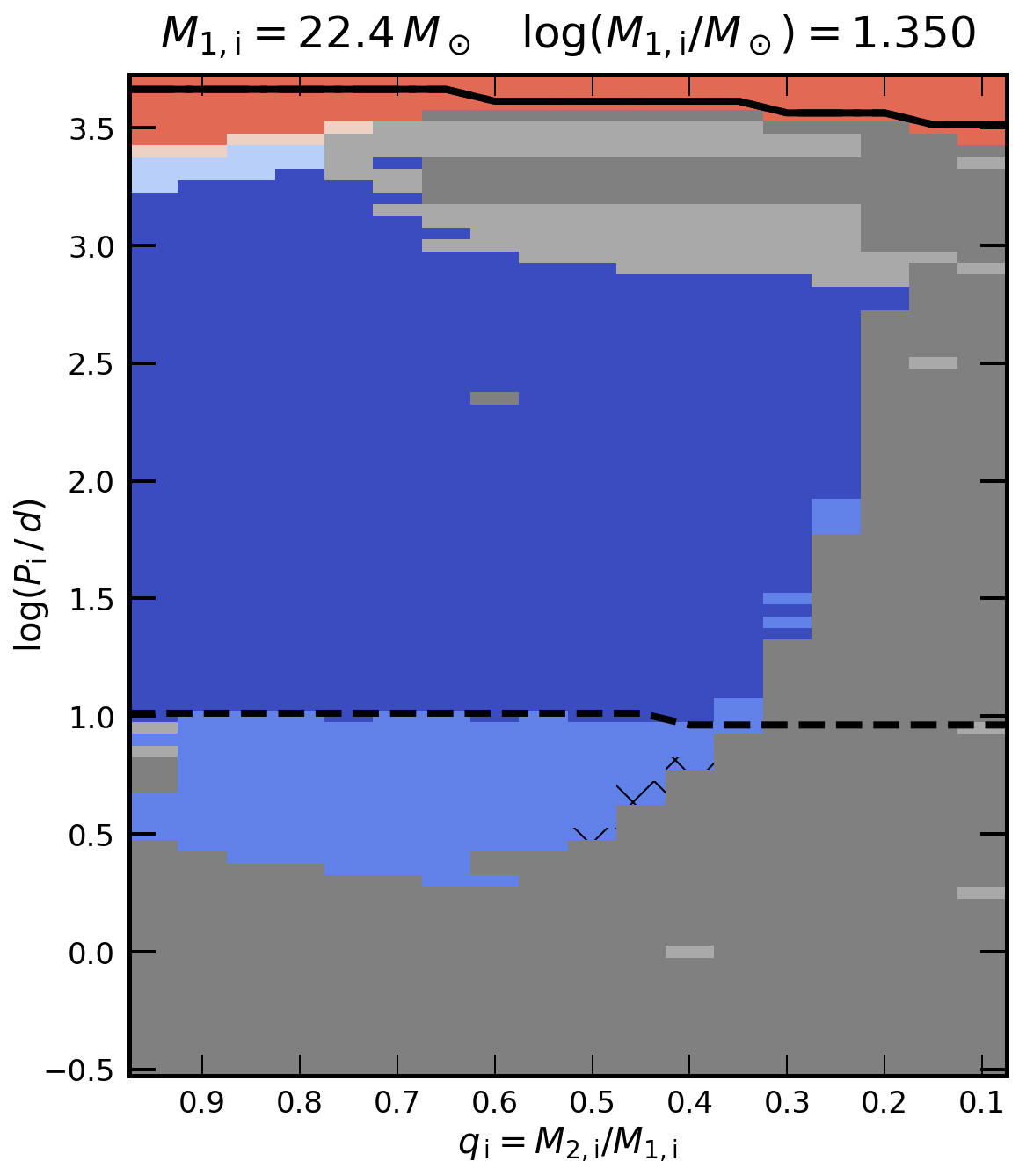}
	\includegraphics[width=0.24\linewidth]{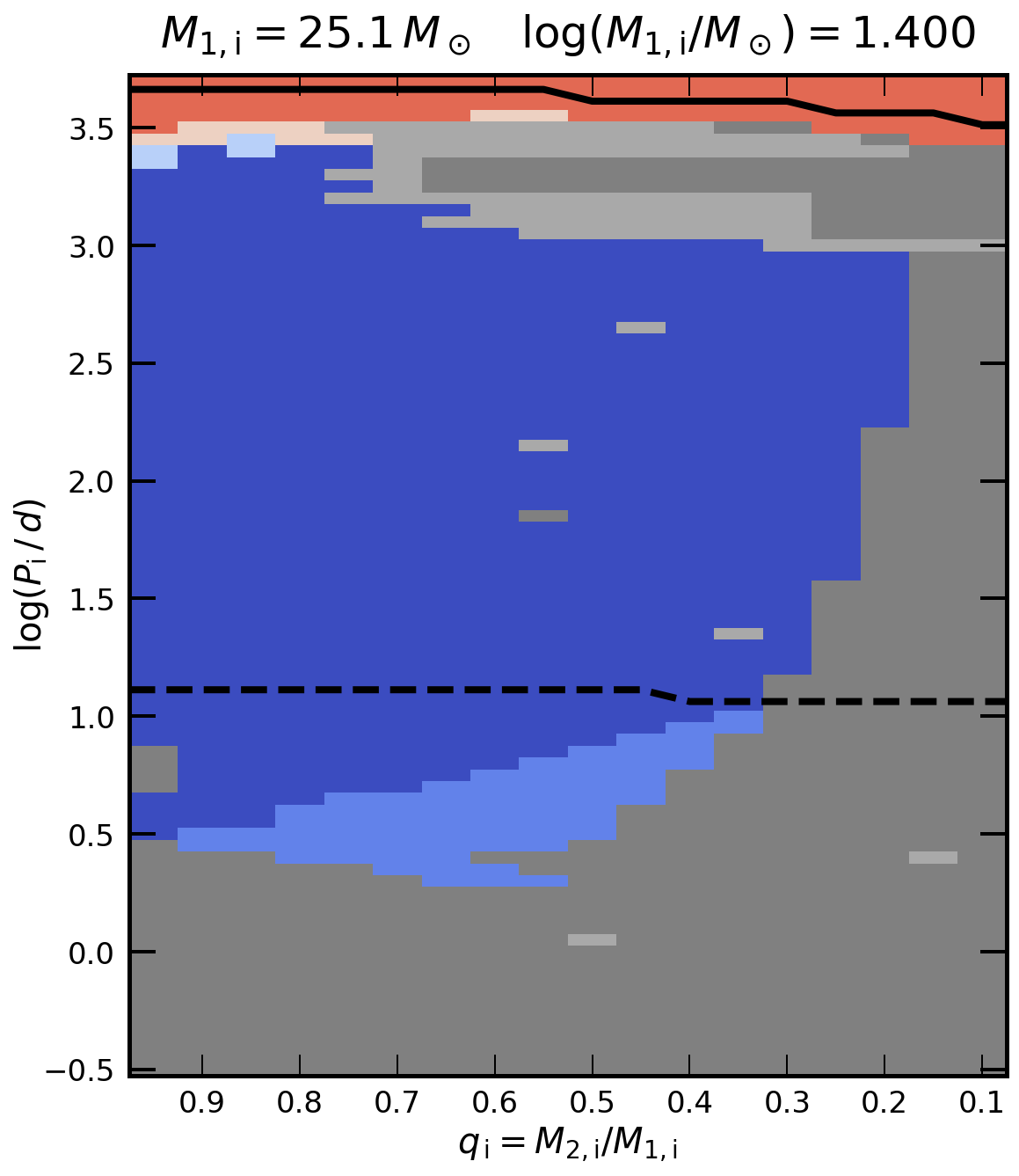}
	\includegraphics[width=0.24\linewidth]{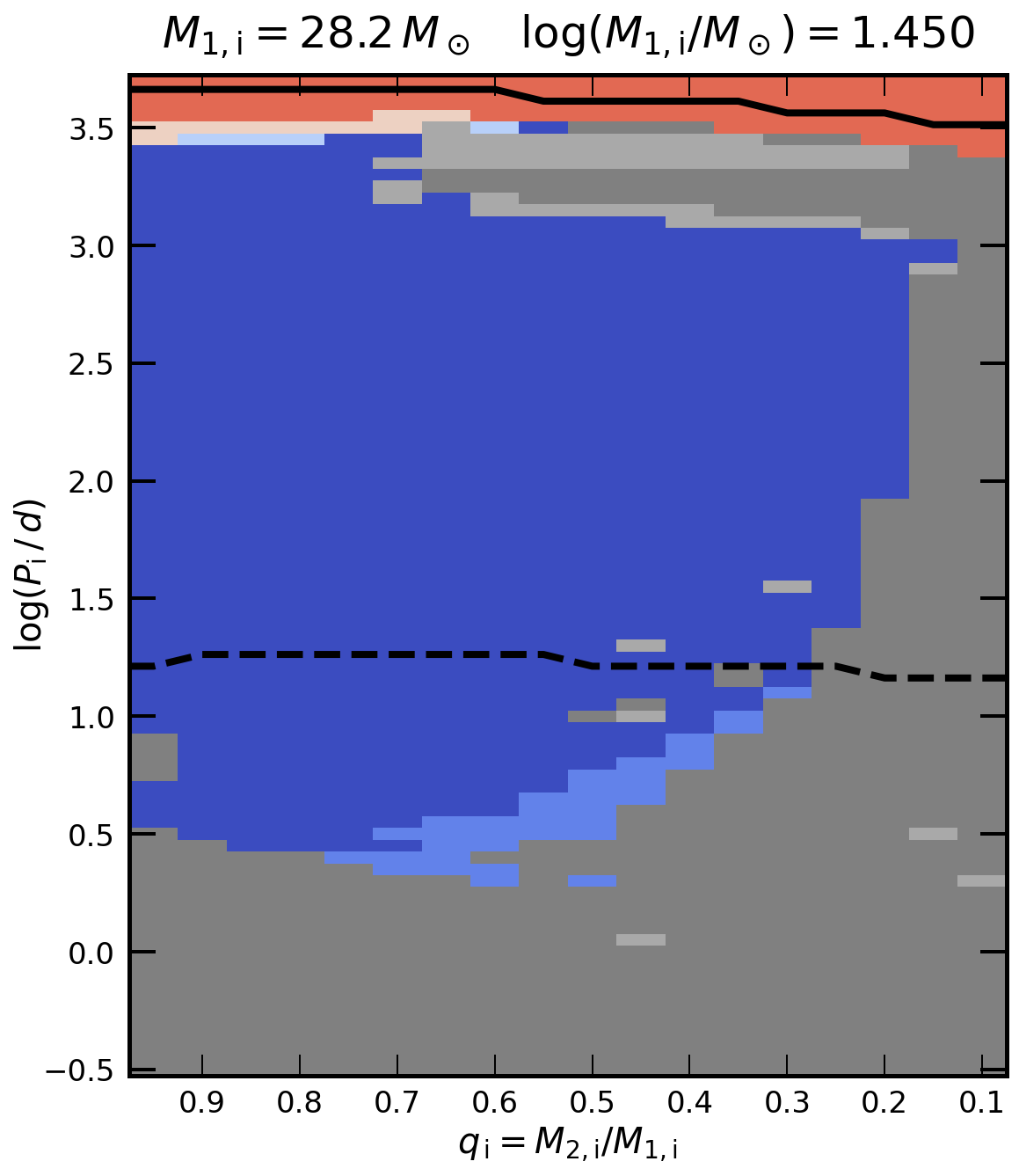}
	\caption{Same as \Fig{fig_Pq}, but for different initial primary masses. The systems in which the secondary is core helium burning are marked with '+' hatching.}
	\label{fig_Pqs1}
\end{figure*}

\begin{figure*}
	\centering
	\includegraphics[width=0.95\linewidth]{figs/Pq_legend.png}
	\includegraphics[width=0.24\linewidth]{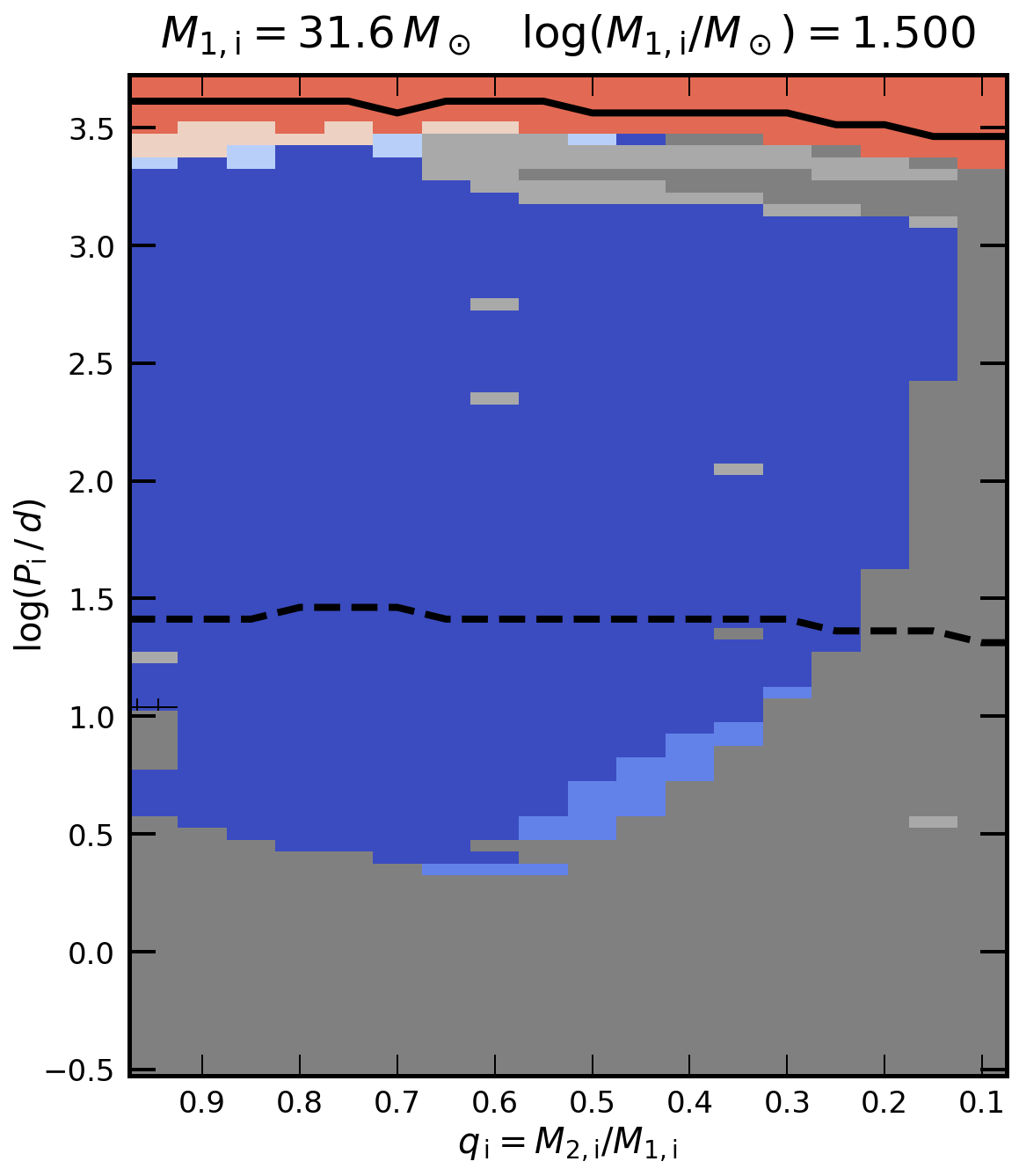}
	\includegraphics[width=0.24\linewidth]{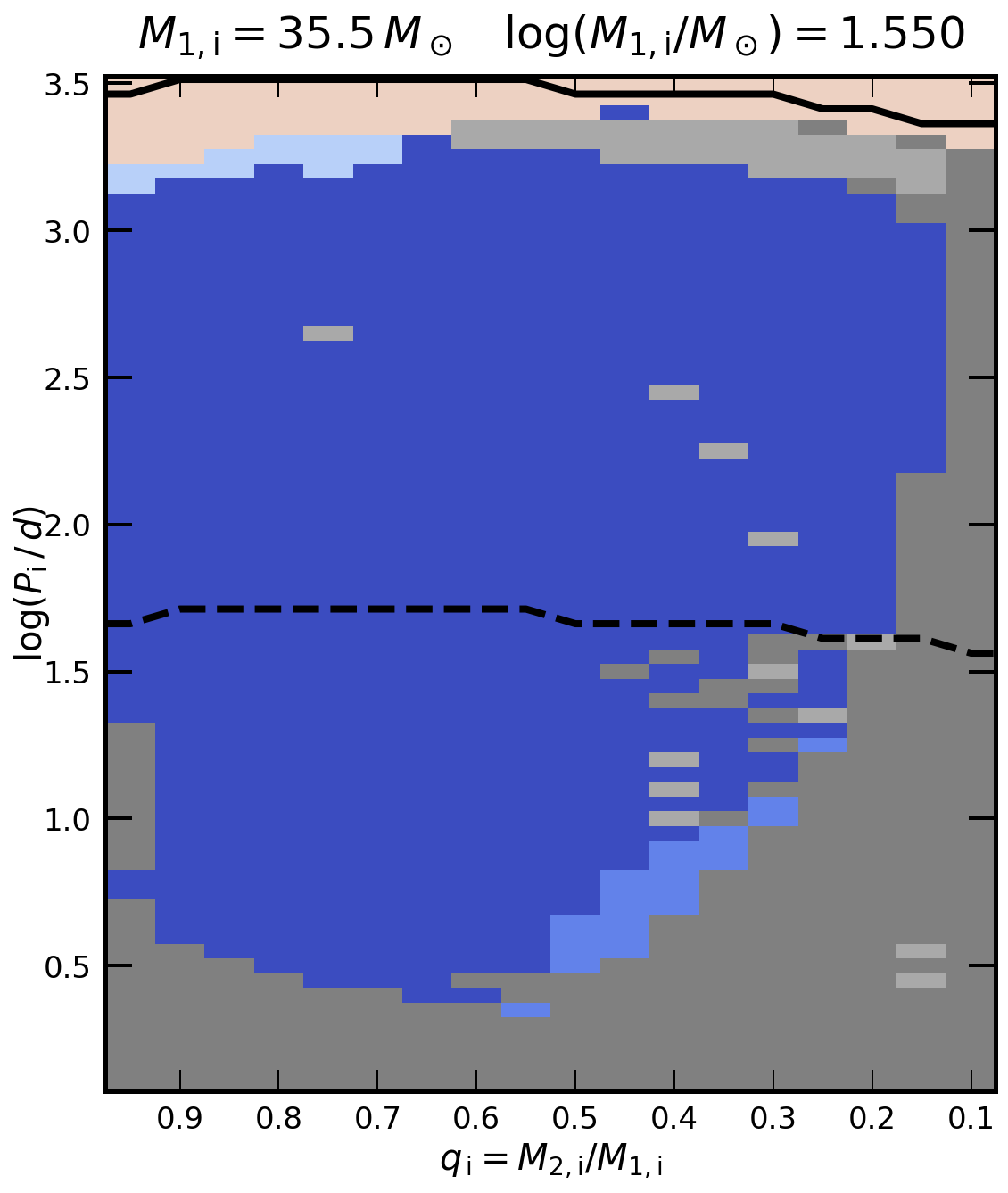}
	\includegraphics[width=0.24\linewidth]{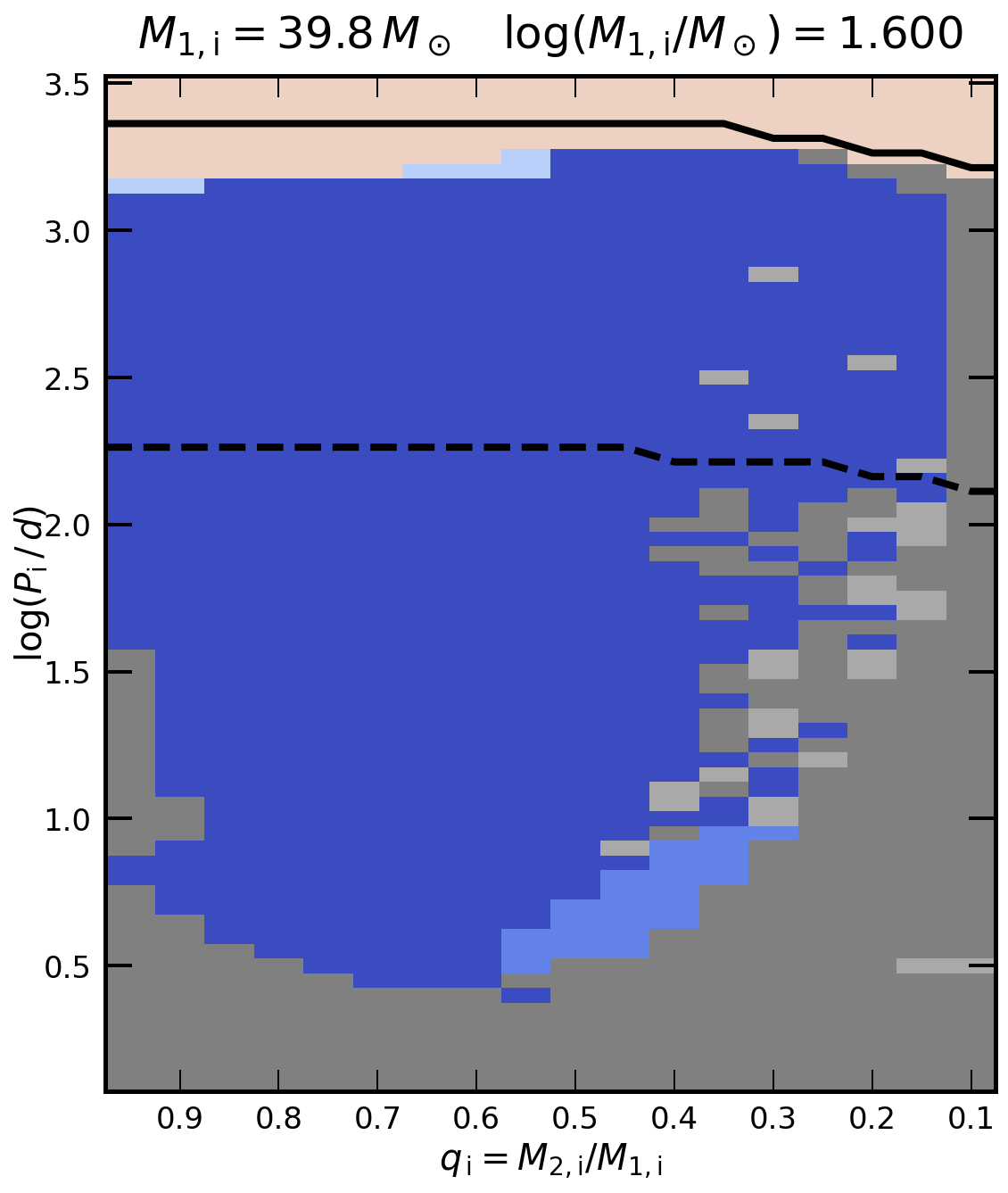}
	\includegraphics[width=0.24\linewidth]{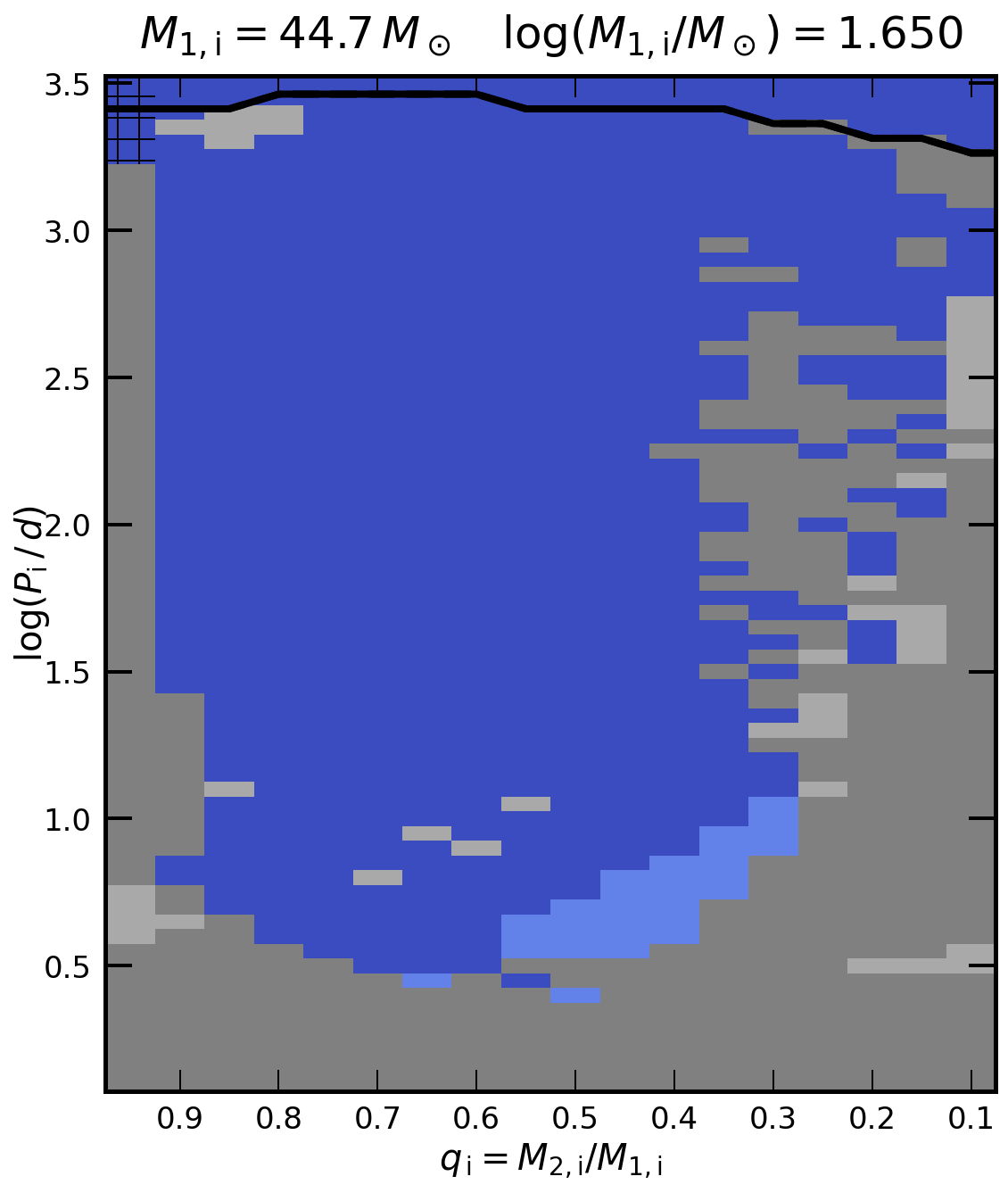}
	\includegraphics[width=0.24\linewidth]{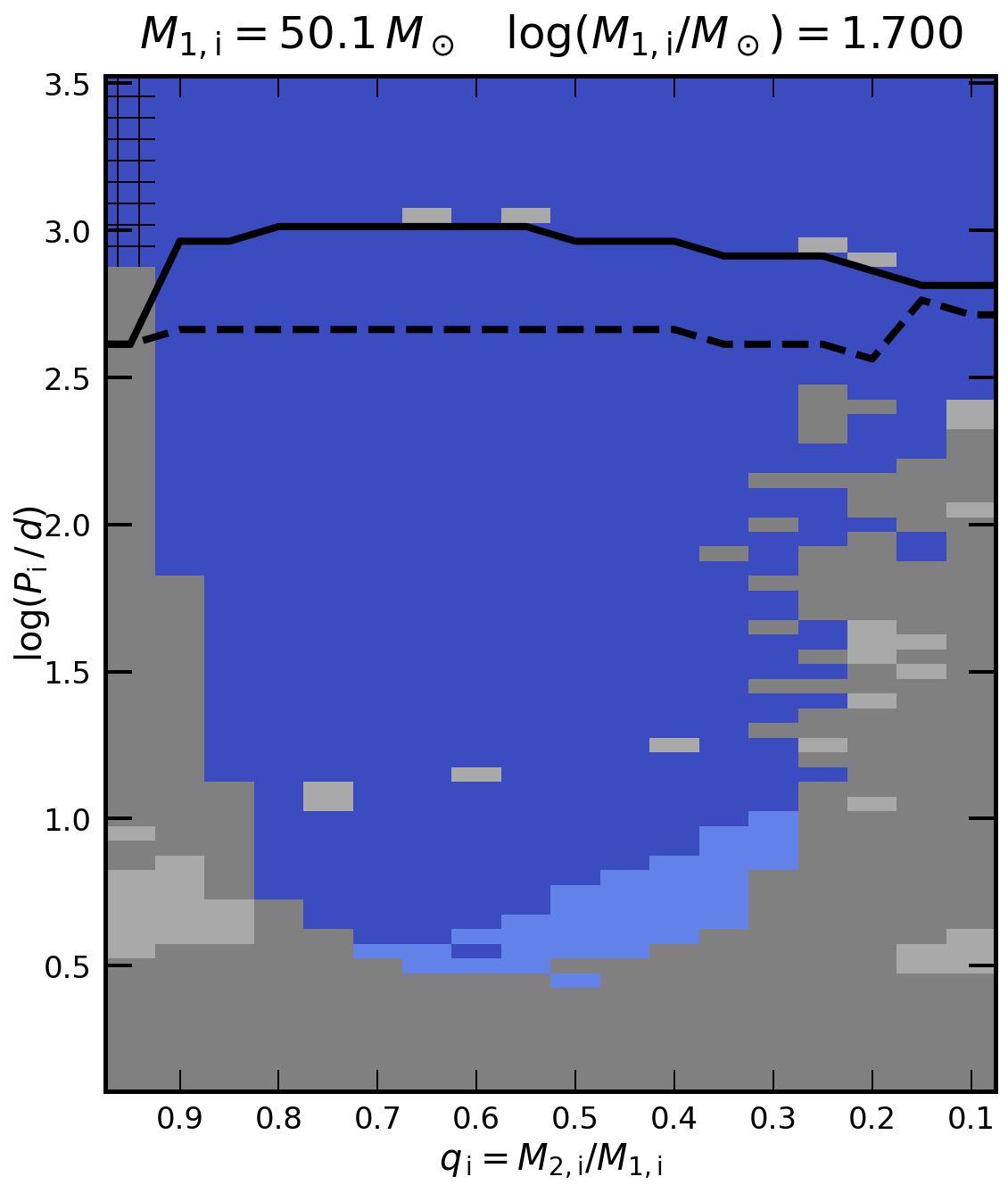}
	\includegraphics[width=0.24\linewidth]{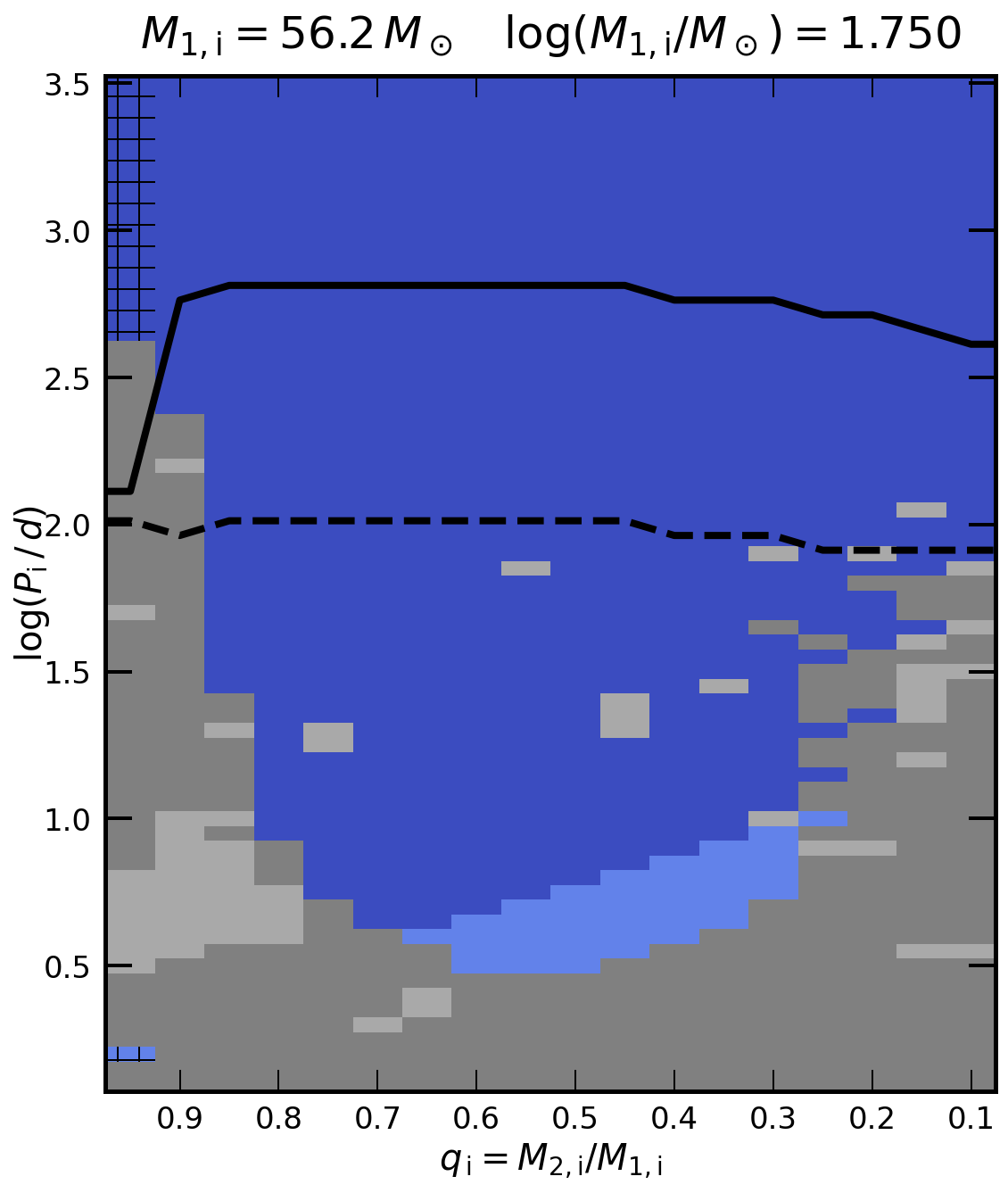}
	\includegraphics[width=0.24\linewidth]{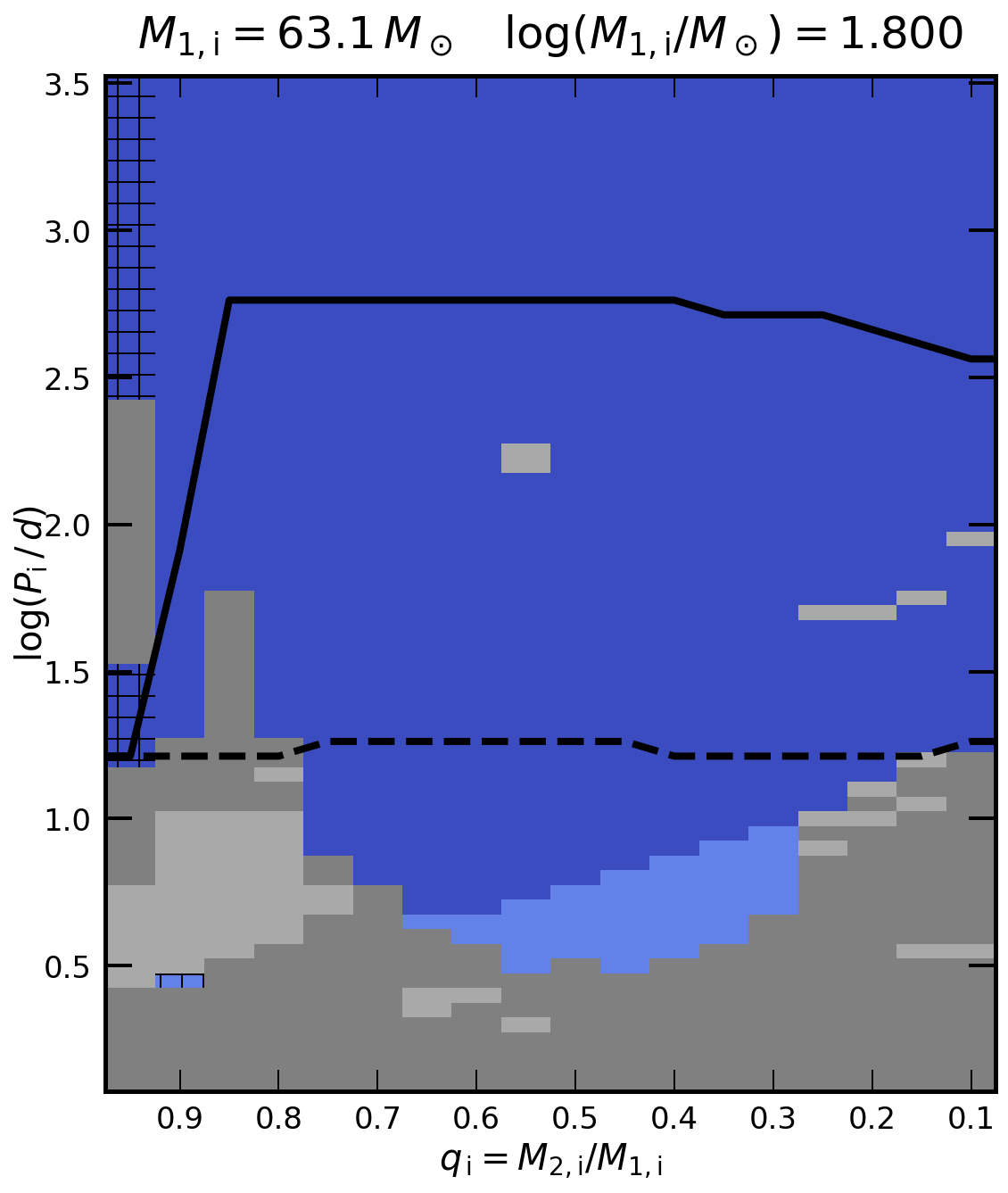}
	\includegraphics[width=0.24\linewidth]{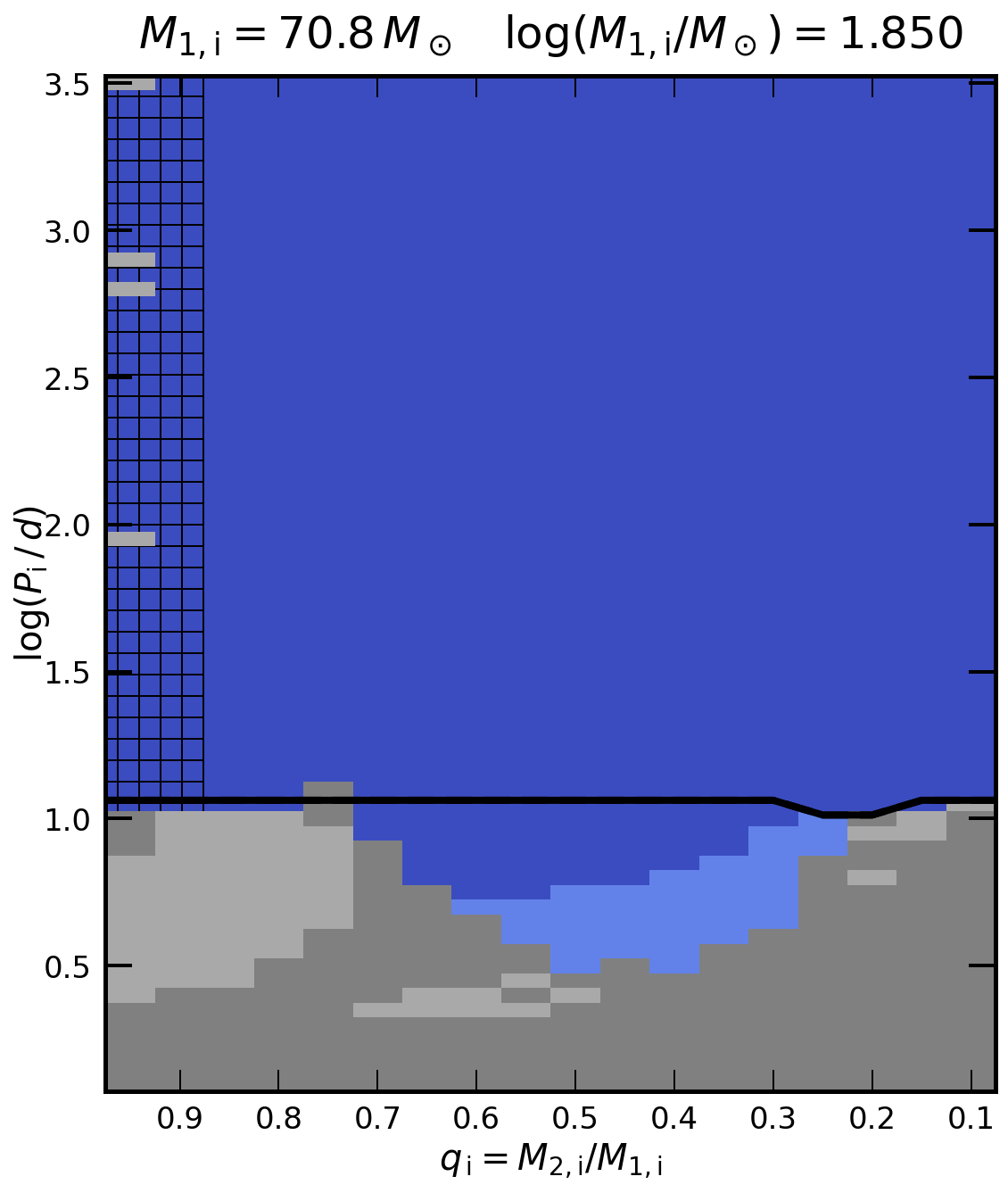}
	\includegraphics[width=0.24\linewidth]{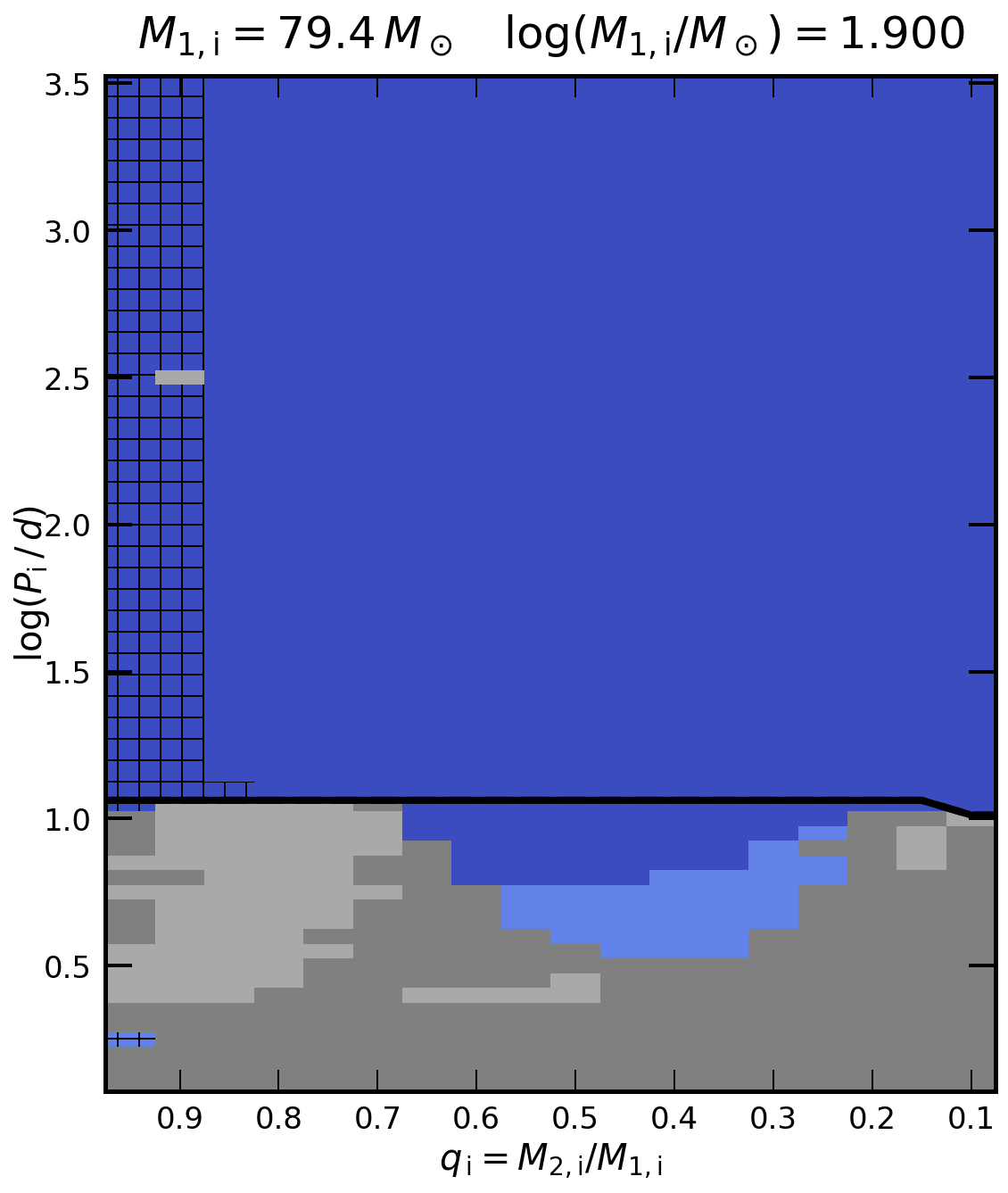}
	\includegraphics[width=0.24\linewidth]{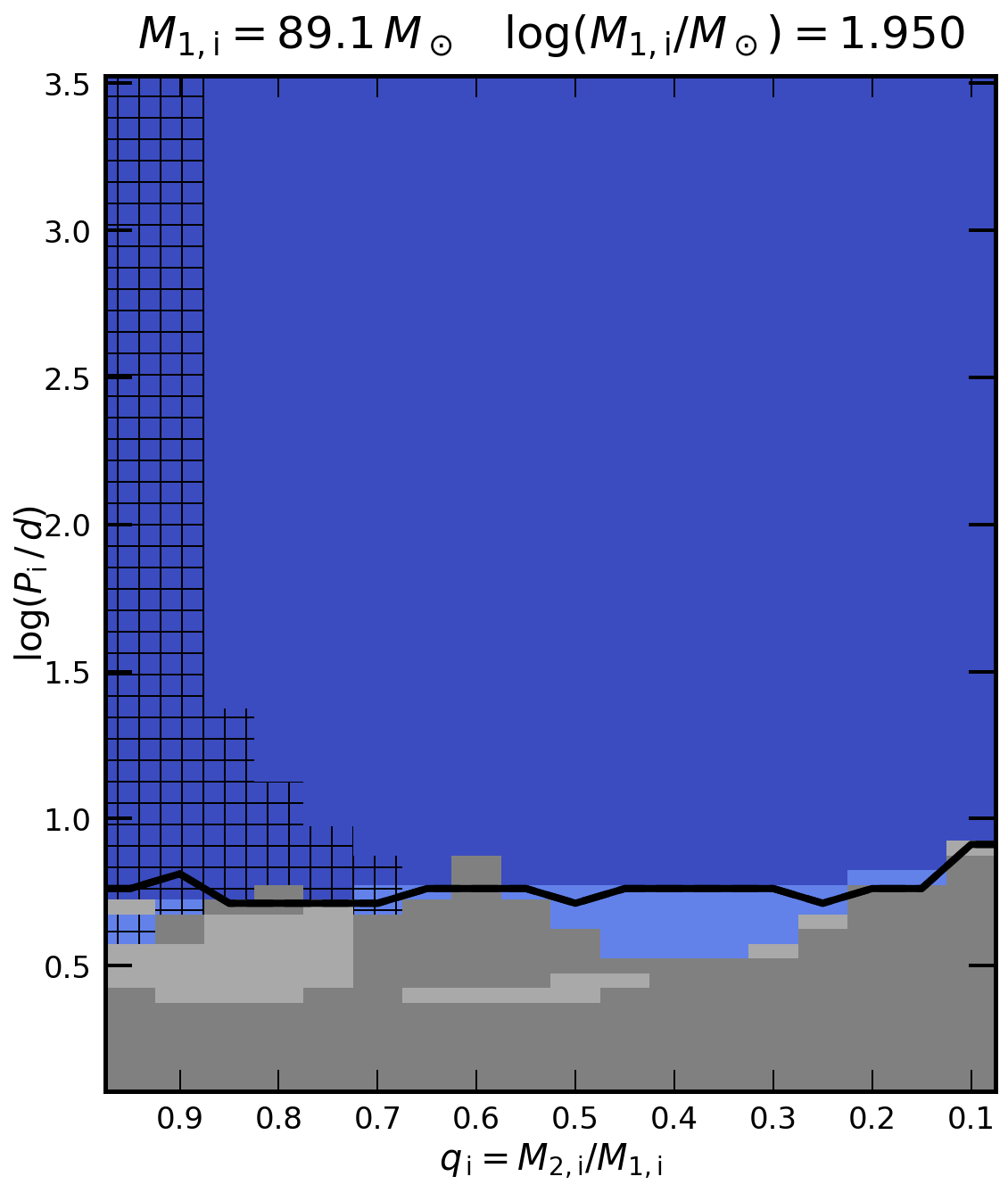}
	\includegraphics[width=0.24\linewidth]{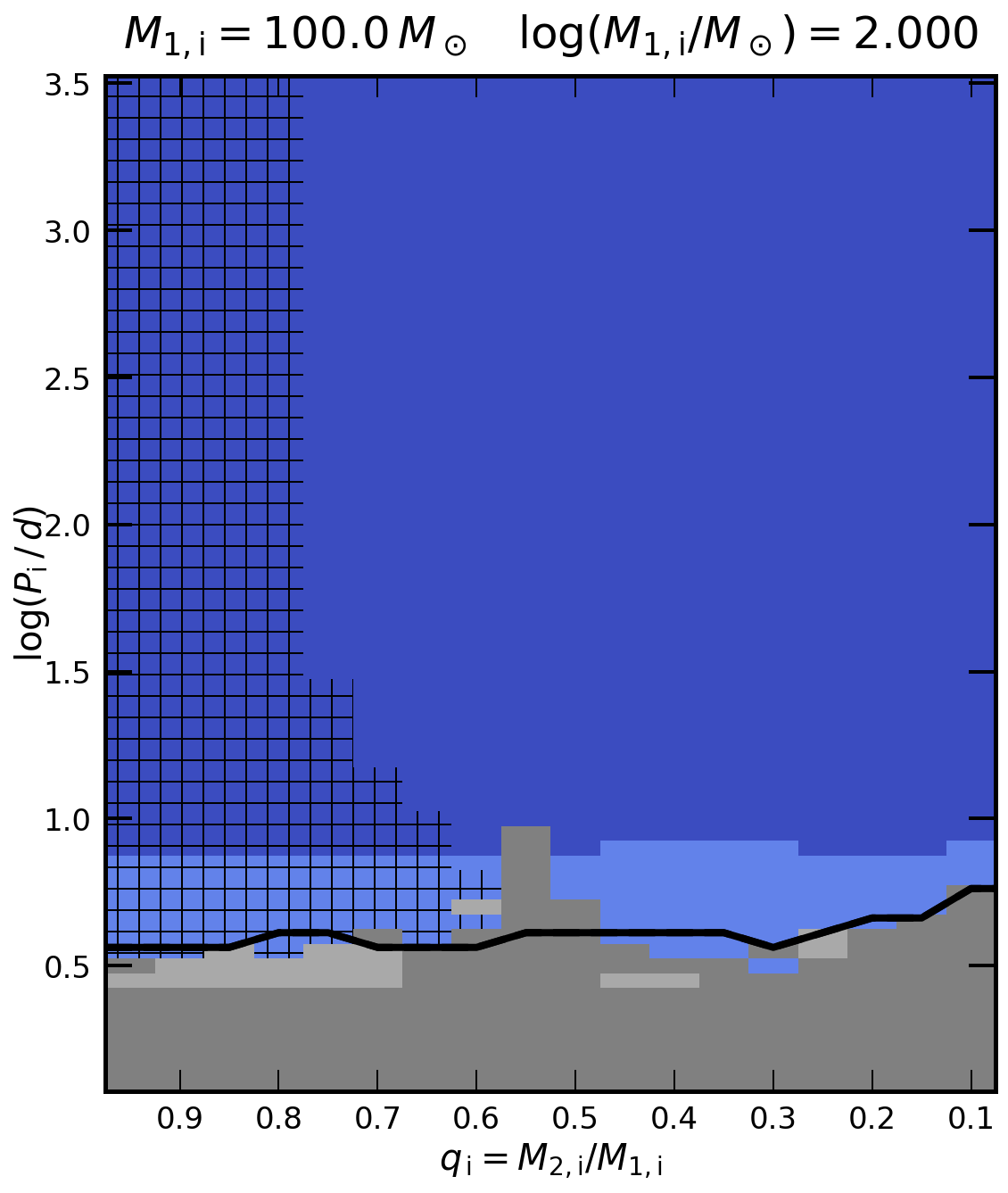}
	\caption{Continued}
	\label{fig_Pqs2}
\end{figure*}

Figures~\ref{fig_3D},~\ref{fig_Pqs1},~\ref{fig_Pqs2} shows the boundaries for different mass transfer cases (Case A, B, C) across the whole initial binary parameter space explored. The boundaries mainly change with the initial primary mass. For example, Case A boundary is shifted higher in initial orbital periods for larger initial primary masses between $5.0\Msun$ and $44.7\Msun$, and decreases afterwards due to strong wind mass loss. A steep slope that the Case A boundary shows at $\lesssim 44.7\Msun$ is due to the envelope inflation \citep{Sanyal2015}. Case B boundary is also shifted higher for higher initial primary masses due to larger size at the tip of the RSG branch, except for the high-mass end where strong stellar wind removes large amounts of envelope during the core hydrogen burning phase. Case B mass transfer does not even occur for $\Mpi \geq 70.8 \Msun$. Case C mass transfer only occurs for $\Mpi \leq 17.8 \Msun$, and the parameter space for the Case C mass transfer generally increases for lower initial primary mass as lower mass helium stars expand more \citet{Yoon2010}.

Note that the binary evolution model grids from the POSYDON project \citep{Fragos2023} and \citet{Henneco2024} are similarly dense and comprehensive in the initial binary parameter space, and also their models follow the evolution of both stellar components simultaneously. However, these models employ a simpler nuclear network than ours (\texttt{approx21} in MESA).

\section{Termination conditions}\label{app_term}
The models are set to terminate when certain conditions are met. Here we list them, including the fraction of models that met each condition. We divide the termination conditions in three groups.

1. When the more evolved star reaches one of the following three conditions, we set the star to a point mass and evolve the companion star seamlessly:
\begin{itemize}
\item Core carbon depletion: The star has completed core carbon burning (the core carbon mass fraction is less than $0.01$).
\item Core helium depletion of a massive helium core: The star has completed core helium burning (central helium mass fraction is less than $10^{-6}$) with a helium core mass of above $13\Msun$. 
\item AGB star or white dwarf: If the primary is in the AGB phase (showing thermal pulses) or expected to become a carbon-oxygen white dwarf (if the mass of the primary is less than $1.1\Msun$ at core helium depletion).
\end{itemize}
We detach the two components infinitely, assuming the binary is disrupted, unless the more evolved star terminated as an AGB/WD. These termination conditions account for $47.9\%$ of all models.

2. Mass transfer can turn unstable, and in such instances, the numerical model and the assumptions adopted are not suited to cover the further evolution of the system, and the simulation is terminated. These cases are listed as follows:
\begin{itemize}
\item Mass transfer rate exceeds $0.1 \mso / \mathrm{yr}$: Such mass transfer is expected to become dynamically unstable and lead to a common envelope phase. This termination condition is also adopted in POSYDON models \citep{Fragos2023}. We find that many wide binaries with $q_\mathrm{i} \geq 0.7$ could successfully undergo stable mass transfer even beyond this rate \citep[see also][]{Ercolino2024}, so we do not apply this termination condition for $q_\mathrm{i} > 0.7$. About 19.5\% of all the models are terminated due to this.
\item Post-main sequence secondary with inverse mass transfer: It is likely to be unstable, leading to a common-envelope phase. About 11.6\% of all the models are flagged as this.
\item L2 overflow: In over-contact systems, if material overflows through the outer Lagrangian point (L2), the system is expected to merge due to significant angular momentum loss. This occurs in about 4.7\% of the models.
\item Roche lobe overflow at ZAMS: If Roche lobe overflow occurs before the primary depletes 3\% of its central hydrogen (to account for initial relaxation of the stars), we expect these models to merge. Approximately 10.8\% of the models fall under this category.
\end{itemize}

3. Finally, some models simply encounter numerical issues that prevent the evolution from going forward. This affects around 5.4\% of all models.

\section{Comprehensive surface abundance predictions}\label{app_abun}

\begin{figure*}
	\centering
	\includegraphics[width=0.95\linewidth]{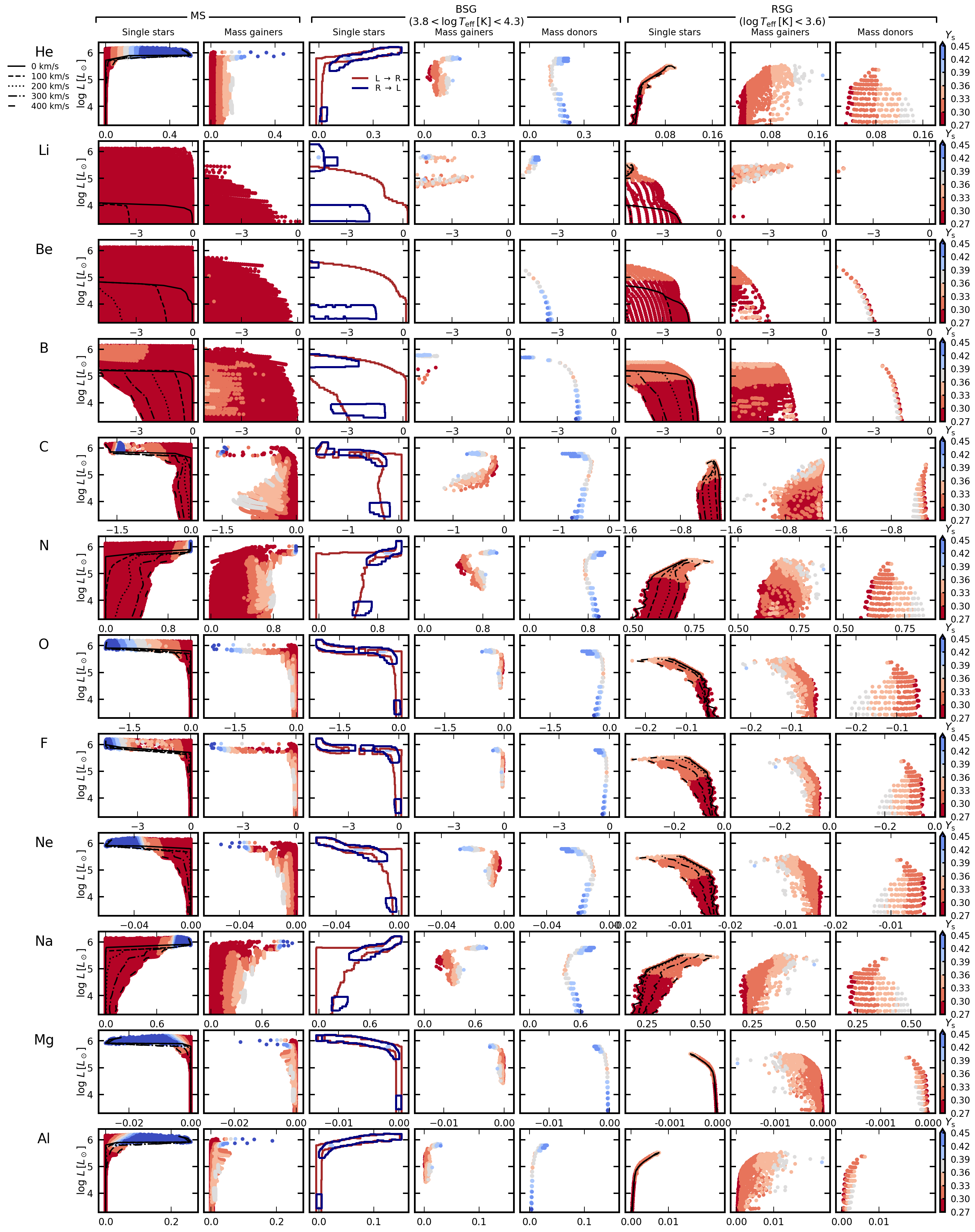}
	\caption{Surface elemental abundances for main sequence stars, BSGs, and RSGs as a function of luminosity. The $x$-axis represents $\log(N/N_\mathrm{i})$, where $N$ is the elemental number fraction and $N_\mathrm{i}$ is the initial value. The elements are indicated in the leftmost panels. Abundances are sampled at 0.1\,Myr intervals for all models within the main sequence band during the core hydrogen burning phase. For BSGs and RSGs, abundances are selected at the midpoint of core helium burning ($\Yc=0.5$). Data points are stacked such that those with higher $\Ys$ appear above those with lower $\Ys$. For single-star models, lines indicate the limits for different initial rotational velocities (the legend in the top-leftmost panel) for main sequence and RSG stars. In the case of BSG single stars, the boundaries that single-star models exhibit while crossing the BSG region left to right (red) and right to left (blue) are shown. 
    }
	\label{fig_abun}
\end{figure*}

\begin{figure*}
	\centering
	\includegraphics[width=0.97\linewidth]{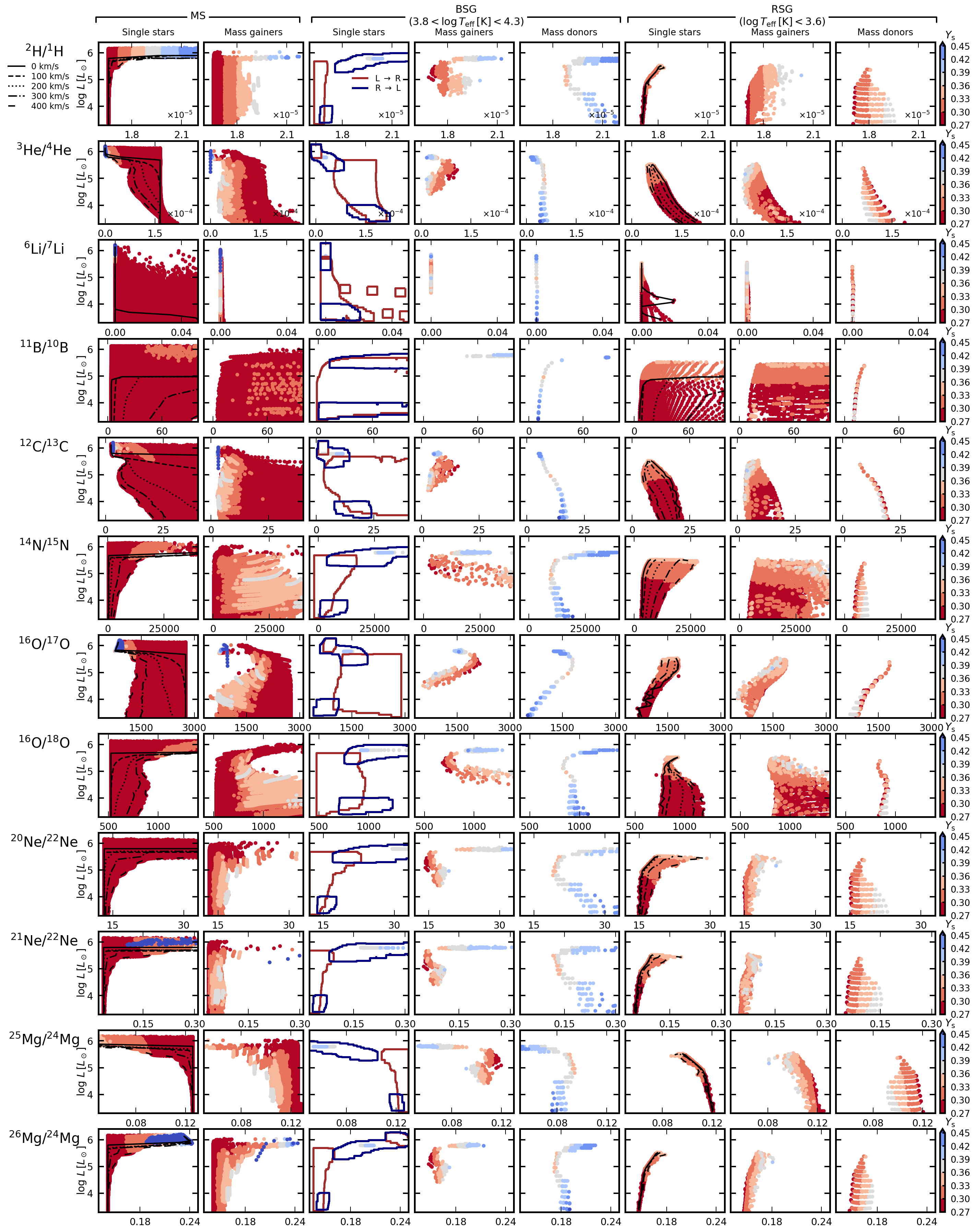}
	\caption{Same as \Fig{fig_abun}, but here the isotopic ratios are presented. The x-axis represents the isotopic ratio, as indicated in the leftmost panels, on a linear scale.}
	\label{fig_iso}
\end{figure*}

\begin{figure*}
	\centering
	\includegraphics[width=0.97\linewidth]{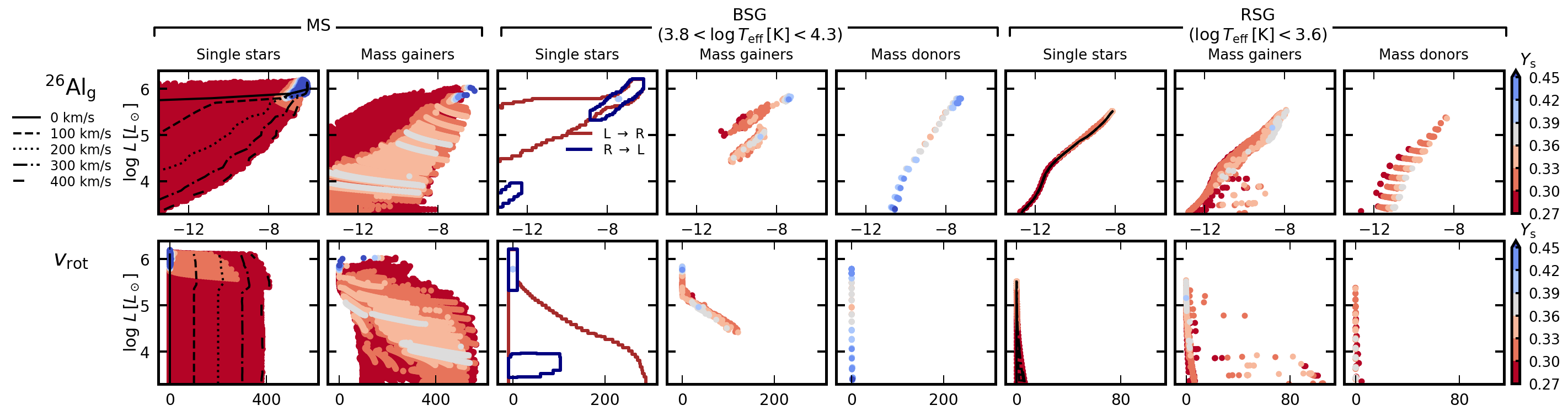}
	\caption{Same as \Fig{fig_abun}, but here the first row shows shows $\log(N)$ for radioactive aluminium-26 and the second row shows the rotational velocity in km/s.}
	\label{fig_al_vrot}
\end{figure*}

Figures~\ref{fig_abun},~\ref{fig_iso},~\ref{fig_al_vrot} present surface abundance predictions from our models during the core hydrogen burning phase (the first two columns) and in the middle of core helium burning (the last six columns) as a function of luminosity (as in \Fig{fig_nni}). These models exhibit surface abundance changes in hydrogen burning products through stellar winds, internal mixing, and binary mass transfer. Hydrogen burning, besides converting hydrogen into helium, changes the abundances of various elements and isotopes through the pp-chains, the CNO-cycle, the Ne-Na cycle, and the Mg-Al cycle, etc. Below, we discuss the surface abundance predictions of a few notable elements and isotopes.

--MS single stars: It was shown that OB stars with relatively low luminosities ($\logLLsun<5.7$ for nitrogen and $\logLLsun<5.1$ for boron) show surface nitrogen enhancement and boron depletion due to rotational mixing (Section~\ref{sec_nni}). In more luminous stars, strong stellar winds play a dominant role, resulting in nitrogen enhancement consistent with CNO-equilibrium and nearly complete boron depletion. The critical luminosity above which wind effect dominate depends on the burning temperatures. For the light elements, the critical luminosity decreases in the order of boron, beryllium, and lithium, following their decreasing burning temperatures of $\sim 7$\,MK, $\sim 5$\,MK, and $\sim 3$\,MK, respectively. Below the luminosity threshold where wind effects dominate, rotational mixing has the most noticeable impact on the surface abundances of lithium, beryllium, boron, carbon, nitrogen, and sodium. Similarly, isotopic abundance ratios such as $^{11}\mathrm{B} / ^{10}\mathrm{B}$ and $^{14}\mathrm{N} / ^{15}\mathrm{N}$, and $^{12}\mathrm{C} / ^{13}\mathrm{C}$ and $^{16}\mathrm{O} / ^{17}\mathrm{O}$ show clear dependencies on the initial rotational velocities.

--MS mass gainers: The surface abundances of these stars are influenced by mass accretion and accretion-induced mixing, different from single stars where only rotational mixing and winds comes into play (Section~\ref{sec_nni}). By comparing helium enrichment between main sequence mass gainers and single stars with relatively low luminosities ($\logLLsun \lesssim 5.7$), we find that mass gainers exhibit noticeable helium enrichment due to the accreting helium-rich material, while single stars do not show any enhancement. Compared to their single star counterparts, mass gainers exhibit more pronounced changes in surface lithium, boron, beryllium, carbon, nitrogen, sodium, as well as in isotopic ratios such as $^{14}\mathrm{N} / ^{15}\mathrm{N}$, $^{16}\mathrm{O} / ^{17}\mathrm{O}$, $^{16}\mathrm{O} / ^{18}\mathrm{O}$.

--BSG single stars: After core hydrogen depletion, single stars expand and cross the Hertzsprung gap on a thermal timescale, during which they appear as a BSG. Surface abundances do not change during this phase. In later evolutionary stages, some single stars also undergo a blueward evolution in the HRD for two luminosity ranges: for $3.5 \lesssim \logLLsun \lesssim 4.0$, stars undergo a blue loop evolution during core helium burning, and for $5.7 \lesssim \logLLsun \lesssim 6.0$, stars lose the envelope due to strong winds (see the blue boundary in Fig.\,\ref{fig_abun}-\ref{fig_iso}). 
The more luminous stars lose a lot of mass while crossing the Hertzsprung gap from left to right and turn left before developing a deep convective envelope, avoiding dredge-up. These stars exhibit surface abundance patterns similar to those crossing the Hertzsprung gap from left to right, except for lithium, beryllium, and boron (the red boundaries in Fig.\,\ref{fig_abun}-\ref{fig_iso}). The less luminous stars have already experienced first dredge-up and exhibit different surface abundances compared to the stars that cross the Hertzsprung gap from left to right (see the blue boundaries for nitrogen and sodium in Fig.\,\ref{fig_abun}).

--BSG mass gainers: Mass gainers in the BSG region exhibit distinctive surface abundances compared to single star counterparts. For example, helium is significantly enhanced in many BSG mass gainers, which are the descendants of helium-rich mass gainers on the main sequence. BSG mass gainers also show stronger nitrogen and sodium enhancement. 

--BSG mass donors: These stars originate from a limited initial binary parameter space (most of mass donors become stripped helium-star or Wolf-Rayet star, not supergiants, see Figs.~\ref{fig_Pqs1} and ~\ref{fig_Pqs2}), namely initially wide interacting binaries. As they are partially stripped and revealed their inner layers, they show noticeable enhancement of elements associated to hydrogen-burning, such as sodium.

--RSG single stars: Once single stars reach the Hayashi line and develop a deep convective envelope, dredge-up can transport the processed material to the surface. This results in surface abundance changes even for non-rotating stars (see the solid lines in RSG single stars panels in Figs.\,\ref{fig_abun}-\ref{fig_iso}) and for stars with weak stellar winds ($\logLLsun \lesssim 4$). The magnitude of surface abundance changes increases with RSG luminosity (and hence mass). This is mainly the result of two effects: more massive stars have stronger mass loss on the main sequence and show significant enhancement/depletion of hydrogen burning products at the surface already before becoming RSGs, 
and they have a larger core-to-envelope mass ratio, resulting in less dilution during the dredge-up. 

--RSG mass gainers: Many of these stars show significant surface helium enrichment, and can appear hotter than single star counterparts (\Fig{fig_HeTeff}). Some RSG mass gainer models show less surface lithium depletion compared to single star counterparts (the lithium depletion factor at $\logLLsun \sim 5.5$ is $\sim 30$ in mass gainers and $\sim 1000$ in single stars, see \Fig{fig_abun}). This can qualitatively help explain some observations, in which some Galactic RSGs are reported to have detectable lithium \citep{Lyubimkov2012, Negueruela2020, Fanelli2022, Negueruela2023}, which is unexpected from canonical single star evolution. We will further investigate this in the future.


--RSG mass donors: 
these stars originate from a limited initial binary parameter space, namely very wide interacting binary systems. They show a wider range of surface helium abundances compared to single stars due to partial stripping. Notably, oxygen is more depleted at the surface in mass donors compared to single stars and mass gainers for $\logLLsun \lesssim 4.5$, as mass stripping reduces the envelope mass, making the dilution due to the first dredge-up less strong. Thus their surfaces are more polluted by CNO-processed matter. These relatively low-mass RSG mass donors also exhibit a large spread in elemental and isotopic abundances (e.g., neon and magnesium). 

\end{appendix}

\end{document}